\documentclass[reprint,superscriptaddress,preprintnumbers,amsmath,amssymb,aps]{revtex4-2}

\usepackage{graphicx}
\usepackage{dcolumn}
\usepackage{bm}
\usepackage[colorlinks=true
,urlcolor=blue
,anchorcolor=blue
,citecolor=blue
,filecolor=blue
,linkcolor=red
,menucolor=blue
,linktocpage=true
,pdfproducer=medialab
,pdfa=true
]{hyperref}
\usepackage[mathlines]{lineno}
\usepackage{xspace}
\usepackage{bbding}
\usepackage{siunitx}
\usepackage{tikz,xcolor}
\usepackage{comment}

\newcommand{\CP}{\ensuremath{CP}\xspace}

\newcommand{\Tsys}{\ensuremath{T_{\rm sys}}\xspace}

\newcommand{\kboltz}{\ensuremath{k_{\rm B}}\xspace}
\newcommand{\ga}{\ensuremath{g_{a\gamma\gamma}}\xspace}
\newcommand{\gaCAST}{\ensuremath{g_{a\gamma\gamma}}^{\scriptscriptstyle {\rm CAST}}\xspace}

\newcommand{\CaB}{\ensuremath{{\rm C}a{\rm B}}\xspace}
\newcommand{\dPEM}{\ensuremath{d_{\rm PEM}}\xspace}

\newcommand{\MW}{{\scriptscriptstyle {\rm MW}}}
\newcommand{\EG}{{\scriptscriptstyle {\rm EG}}}
\newcommand{\tU}{t_{\scriptscriptstyle {\rm U}}}

\begin{document}

\preprint{APS/123-QED, CERN-TH-2023-041}

\title{Search for a dark-matter induced Cosmic Axion Background with ADMX}
\author{T. Nitta}\email[Correspondence to: ]{tnitta@icepp.s.u-tokyo.ac.jp}
\affiliation{University of Washington, Seattle, Washington 98195, USA}
\affiliation{currently International Center for Elementary Particle Physics (ICEPP), The University of Tokyo, Tokyo 113-0033, Japan}

\author{T. Braine}
\author{N. Du}
\author{M. Guzzetti}
\author{C. Hanretty}
\author{G. Leum}
\author{L. J Rosenberg}
\author{G. Rybka}
\author{J. Sinnis}
  \affiliation{University of Washington, Seattle, Washington 98195, USA}

\author{John Clarke}
\author{I. Siddiqi}
  \affiliation{University of California, Berkeley, California 94720, USA}


\author{M. H. Awida} 
\author{A. S. Chou} 
\author{M. Hollister}
  \affiliation{Fermi National Accelerator Laboratory, Batavia, Illinois 60510, USA}

\author{S. Knirck}
\author{A. Sonnenschein} 
\author{W. Wester} 
  \affiliation{Fermi National Accelerator Laboratory, Batavia, Illinois 60510, USA}
%
\author{J.~R.~Gleason}
\author{A. T. Hipp}
\author{P. Sikivie}
\author{N. S. Sullivan}
\author{D. B. Tanner}
  \affiliation{University of Florida, Gainesville, Florida 32611, USA}


\author{R. Khatiwada}
  \affiliation{Illinois Institute of Technology, Chicago, Illinois 60616, USA}
  \affiliation{Fermi National Accelerator Laboratory, Batavia, Illinois 60510, USA}

\author{G. Carosi}
\author{N. Robertson}
  \affiliation{Lawrence Livermore National Laboratory, Livermore, California 94550, USA}

\author{L. D. Duffy}
  \affiliation{Los Alamos National Laboratory, Los Alamos, New Mexico 87545, USA}

\author{C. Boutan}
\author{E. Lentz}
\author{N. S.~Oblath}
\author{M. S. Taubman}
\author{J.~Yang}%
  \affiliation{Pacific Northwest National Laboratory, Richland, Washington 99354, USA}

\author{E. J. Daw}
\author{M. G. Perry}
  \affiliation{University of Sheffield, Sheffield S10 2TN, UK}

\author{C. Bartram}
  \affiliation{Stanford Linear Accelerator Center, Menlo Park, California 94025, USA}

\author{J. H. Buckley}
\author{C. Gaikwad}
\author{J. Hoffman}
\author{K. W. Murch}
  \affiliation{Washington University, St. Louis, Missouri 63130, USA}

\author{M. Goryachev}
\author{E. Hartman}
  \affiliation{University of Western Australia, Perth, Western Australia 6009, Australia}

\author{B. T. McAllister}
  \affiliation{University of Western Australia, Perth, Western Australia 6009, Australia}
  \affiliation{Swinburne University of Technology, Melbourne, Victoria 3122, Australia}
\author{A. Quiskamp}
\author{C. Thomson}
\author{M. E. Tobar}

  \affiliation{University of Western Australia, Perth, Western Australia 6009, Australia}

\collaboration{ADMX Collaboration}\noaffiliation

\author{J. A. Dror}
  \affiliation{Santa Cruz Institute for Particle Physics and Department of Physics, University of California, 1156 High St, Santa Cruz, CA 95060, USA}
\author{H. Murayama}
  \affiliation{University of California, Berkeley, California 94720, USA}
  \affiliation{Lawrence Berkeley National Laboratory, Berkeley, California 94720, USA}
  \affiliation{Kavli Institute for the Physics and Mathematics of the Universe (WPI), The University of Tokyo, Kashiwa 277-8583, Japan}
\author{N. L. Rodd}
  \affiliation{Theoretical Physics Department, CERN, 1 Esplanade des Particules, CH-1211 Geneva 23, Switzerland}


\begin{abstract}
We report the first result of a direct search for a Cosmic {\it axion} Background (\CaB) -- a relativistic background of axions that is not dark matter -- performed with the axion haloscope, the Axion Dark Matter eXperiment (ADMX).
Conventional haloscope analyses search for a signal with a narrow bandwidth, as predicted for dark matter, whereas the \CaB will be broad.
We introduce a novel analysis strategy, which searches for a \CaB induced daily modulation in the power measured by the haloscope.
Using this, we repurpose data collected to search for dark matter to set a limit on the axion photon coupling of a \CaB originating from dark matter cascade decay via a mediator in the 800--995~MHz frequency range.
We find that the present sensitivity is limited by fluctuations in the cavity readout as the instrument scans across dark matter masses.
Nevertheless, we suggest that these challenges can be surmounted using superconducting qubits as single photon counters, and allow ADMX to operate as a telescope searching for axions emerging from the decay of dark matter.
The daily modulation analysis technique we introduce can be deployed for various broadband RF signals, such as other forms of a \CaB or even high-frequency gravitational waves.
\end{abstract}

\maketitle

Axions, originally motivated by their simple solution to the strong \CP problem~\cite{PhysRevD.16.1791,PhysRevLett.40.223,Peccei1977June,PhysRevLett.40.279}, have since been accepted more broadly as a compelling extension of the Standard Model.
Most searches for axions that are relics of the early Universe assume they make up a local non-relativistic fluid, which is characteristic of dark matter~\cite{RevModPhys.93.015004,Hook:2018dlk,adams2023axion}.
However, a local axion energy density could take on other forms.
Axions could be produced in the early Universe thermally, via parametric resonance, the decay of topological defects, or alternatively could emerge in the late Universe from the decay of another dark matter candidate~\cite{PhysRevD.103.115004}.
Collectively, the relativistic abundance of such axions would form a Cosmic $axion$ Background (\CaB): an axion analog of the photons in the cosmic microwave background.
The \CaB, if produced in the early Universe, would constitute a form of dark radiation and therefore a contribution to $\Delta N_{\rm eff}$, which CMB-S4 will be well placed to detect~\cite{CMB-S4:2016ple} (and in fact, there may even be a mild hint for from the Hubble tension~\cite{PhysRevLett.117.171301,Verde2019}).
In the late Universe, the \CaB will constitute a local axion energy density, analogous to axion dark matter, and therefore can be searched for with axion haloscopes~\cite{PhysRevLett.51.1415,PhysRevD.99.052012,Gelmini:2020kcu,arvanitaki2014resonantly,BREAD:2021tpx,Aybas2021a,Wu:2019AxionDM,Brouwer:2022bwo,Salemi:2021gck,Backes2021,Baryakhtar:2018doz,Brun:2019lyf,Gramolin2021,Irastorza:2011gs,2022APS..APRS10006E}.
The principal difference, however, is that whereas dark matter is narrow spectrally, the \CaB can be comparatively very broad.
In a $\sim$$1~{\rm GHz}$ frequency range, dark matter has a bandwidth of ${\cal O}(1~{\rm kHz})$, whereas the \CaB bandwidth could be ${\cal O}(100~{\rm MHz})$ or larger.

In this work, we perform the first direct search for the \CaB with the Axion Dark Matter eXperiment (ADMX)~\cite{PhysRevD.64.092003,PhysRevD.69.011101,Asztalos_2002,PhysRevLett.104.041301,PhysRevLett.120.151301,PhysRevLett.124.101303,PhysRevLett.127.261803}, and therefore provide a direct search for additional dark radiation.
To do so, we focus on one specific form the \CaB could take: relativistic axions, $a$, emerging from the decay of dark matter particles, $\chi$.
The search can therefore be considered as an extension of the dark matter indirect detection program, with axions acting as final states.
As we will review, this signal gives rise to power in a resonant cavity that modulates over the course of a day.
Accordingly, we introduce an analysis strategy to directly search for modulating power in an existing ADMX dataset, and in this way construct the first analysis for the \CaB~\footnote{For an alternative detection scheme which exploits the upcoming Square Kilometer Array to search for photon signals originating from the conversion of relativistic axions in astrophysical magnetic fields, see Ref.~\cite{kar2022searching}}.

\vspace{0.2cm}
\noindent {\bf ADMX as a telescope for dark matter decay.}
%
We begin by outlining the details of the model we consider and the power it could generate in ADMX.
We search for the axions arising from the decay of dark matter.
The simplest realization of such a scenario would appear to be a two-body decay, $\chi \to aa$.
However, as discussed in Ref.~\cite{PhysRevD.103.115004}, axions are bosons so that such a decay can be enormously Bose enhanced, which would lead to a runaway decay and rapidly deplete the dark matter.
The runaway would not occur if the decay is slightly modified, for instance by the inclusion of an intermediate state $\varphi$ in a cascade decay.
For this reason, we focus on axions arising from the decay channel $\chi \to \varphi\varphi \to aaaa$.
In general, the spectrum of axions produced in the decays will depend on $m_{\rm DM}$, $m_\varphi$, and $m_a$.
As we focus on relativistic axions, we always assume $m_a$ is negligible.
If we further assume $m_\varphi \ll m_{\rm DM}$, then the spectrum of axions depends on two parameters: $m_{\rm DM}$ and the dark matter lifetime, $\tau$ (see, e.g., Refs.~\cite{Elor:2015tva,Elor:2015bho}).
We emphasize that our focus here is on studying scenarios that realize these kinematics; there will be a variety of models in which such a cascade scenario could arise.
In the supplementary material we provide an explicit example that realizes these kinematics: a cubic dark matter-mediator interaction and a quadratic axion derivative coupling to the mediator.
While there is no dedicated experimental search of this scenario, there are constraints on the lifetime for dark matter to decay into a relativistic species, for instance, Dark Energy Survey (DES) measurements require at 95\% C.L. $\tau>25~{\rm Gyr} \sim 1.8\,\tU$~\cite{PhysRevD.103.123528}, where $\tU$ is the present age of the universe~\footnote{To derive this value from Ref.~\cite{PhysRevD.103.123528}, we set the ratio of the amount of dark matter converted into dark radiation divided by the amount of dark matter today, which we approximate by $(1-e^{-\tU/\tau})/e^{-\tU/\tau}$, equal to their 95\% C.L. limit on $\zeta$, marginalized over all nuisance parameters (0.72)}. We expect this to reasonably approximate the bound on the lifetime when the fraction of dark matter which decayed is small and the lifetime of any intermediate particle is negligible compared to the age of the universe..

As detailed in the supplementary material, the resulting \CaB from the dark matter decays will generate the following differential power in a resonant cavity,
\begin{equation}\begin{aligned}
\frac{dP_a}{d\omega}(\omega,\alpha)&=\frac{\rho_c \ga^2B^2VC\beta}{8Q((\omega-\omega_0)^2 + (\omega_0/2Q)^2)}\frac{\omega_0^3}{\omega^3}\\
&\times\left[\Omega_a^\MW(\omega)\int dzd\phi \mathcal{D}_\nu(-z) K(\omega,\alpha)\right.\\
&\hspace{1.15cm}+\left.\frac{1}{2}\Omega_a^\EG(\omega)\int dzd\phi K(\omega,\alpha)\right]\!.
\label{eq:dPdw}
\end{aligned}\end{equation}
The power depends on properties of the cavity and of the axion.
For the cavity, $B$ is the external magnetic field, $V$ is its volume, $Q$ is the quality factor, $\beta$ is the coupling between the cavity and antenna, $C$ is the cavity form factor for the TM$_{010}$ mode~\cite{PhysRevLett.127.261803}, and $\omega_0$ is the resonant angular frequency of the same mode.
For the axion, we have the axion energy $\omega$ and the expressions in parentheses detail the flux of axions arising from decays within the Milky Way (MW) and extragalactic (EG) decays.
The full expressions for the flux are given in the supplementary material, but briefly, $\Omega_a(\omega)$ describes the spectrum of axions and depends on the lifetime $\tau$, vanishing as $\tau \to \infty$, whereas the angular distributions are controlled by the integrands. 
Specifically, we integrate over the full sky using spherical coordinates $\phi$ and $\theta = {\rm arccos}\,z$.
The parameter ${\cal D}_\nu(z)$ describes the angular dependence of decays in the MW, which occur primarily at the Galactic Center, and there is no analog for the EG flux as the flux of axions is essentially isotropic.
The normalization factor, $\rho_c \simeq 4.8~{\rm keV/cm}^3$, denotes the critical density.

\begin{figure}[!t]
\centering
\includegraphics[width=\linewidth]{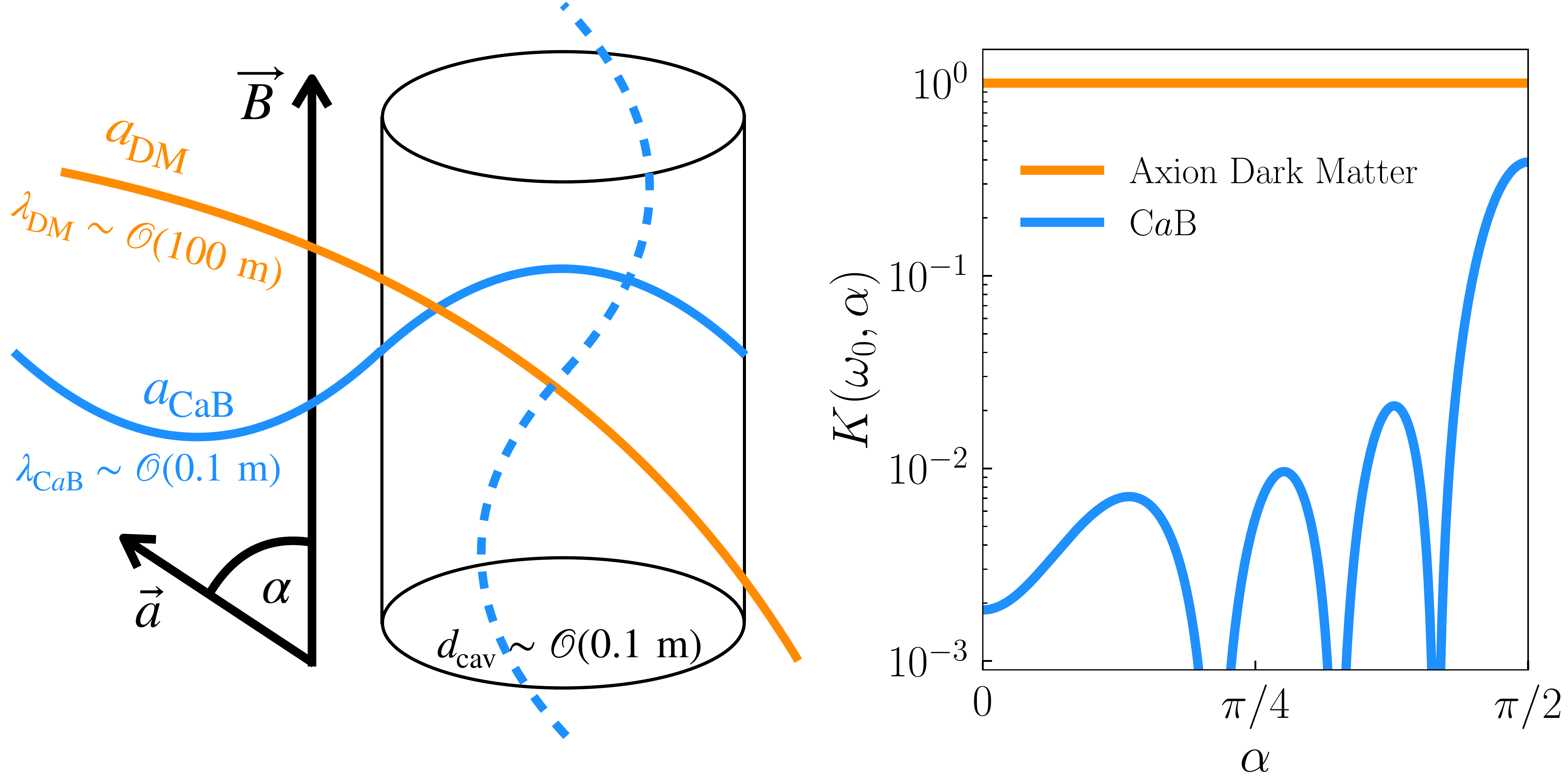}
\vspace{-0.5cm}
\caption{{\it Left:} a schematic depiction of the spatial coherent length of a dark matter ($\lambda_{\rm DM}\sim {\cal O}(100~{\rm m})$, orange) or \CaB ($\lambda_{\CaB}\sim {\cal O}(0.1~{\rm m})$, light blue) signal over the ${\rm TM}_{010}$ mode of a cylindrical cavity ($d_{\rm cav}\sim {\cal O}(0.1~{\rm m})$, black cylinder).
The dark matter signal is coherent across the cavity, and therefore independent of the incident direction, whereas the \CaB is not, and therefore the dashed and solid curves can give rise to significantly different signals.
{\it Right:} a quantification of this effect using the expression in Eq.~\eqref{eq:K}, where $\alpha$ is the angle between the incident axion and the cavity magnetic field, $R=0.2$~m, and $L=1$~m.
The power of photons converted from the \CaB is maximized at $\alpha=\pi/2$, when the cavity magnetic field is perpendicular to the incident axions, and suppressed when $\alpha=0$.
}
\vspace{-0.5cm}
\label{fig:CaBff}
\end{figure}

The final ingredient in Eq.~\eqref{eq:dPdw}, $K(\omega,\alpha)$, generates the daily modulation, and describes the fact that the relativistic axion is not spatially coherent across the instrument.
As the ADMX cavity is not a sphere, the coherence across the cavity will depend on $\alpha$, the angle between the axion, $\hat{\bf k}$, and the magnetic field, $\hat{\bf z}$.
For an ideal cylindrical cavity of height $L$ and radius $R$, we can compute~\cite{PhysRevD.103.115004}
\begin{equation}
K(\omega,\alpha) = \left[ \frac{\sin(\omega L \cos \alpha)}{\omega L \cos \alpha} \frac{J_0(\omega R \sin \alpha)}{1-(\omega R \sin \alpha / j_{01})^2}\right]^2\!,
\label{eq:K}
\end{equation}
where $J_0$ is a Bessel function of the first kind, and $j_{01}$ is that function's first zero.
A depiction of why this factor arises for the \CaB, but not dark matter, is given in Fig.~\ref{fig:CaBff}.
Axion dark matter (orange) has a significantly larger de Broglie wavelength than its Compton wavelength.
Consequently, it sources a spatially coherent effect across the cavity.
This is not the case for the \CaB (light blue), which has a de Broglie wavelength comparable to its Compton wavelength and therefore can oscillate spatially across the cavity.
The effect of the resulting interference is described by $K(\omega,\alpha)$.
While this factor suppresses the overall power ($K(\omega,\alpha) \leq 1$), the way in which it does so is critically dependent on the incident angle of the wave -- for instance, a wave incident down the height of the cavity will oscillate many more times than one incident radially, given that $L \gg R$.
Combining this effect with the fact that the axions originate from dark matter decays in the MW arise preferentially from the Galactic Center and that cavity magnetic field orientation will rotate with the Earth throughout the day, results in a unique daily modulation signal that we will exploit to search for the \CaB~\footnote{We emphasize that the daily modulation signal we describe arises solely from the interaction of the \CaB wave with the cavity.
This can be contrasted with other possible modulating signals, such as that proposed for Axion Quark Nugget dark matter, which generates a relativistic axion signal where the amplitude of the signal oscillates throughout the day, see Refs.~\cite{Fischer:2018niu,Liang:2019lya,Budker:2019zka}.}.

\vspace{0.2cm}
\noindent {\bf A daily modulation analysis.}
%
Having defined the signal, we now outline the analysis we intend to perform on existing ADMX data, which will fundamentally make use of the daily modulation emerging from $K(\omega,\alpha)$.
Briefly, ADMX uses an axion haloscope designed to search for axions that could constitute the local dark matter halo using a cold resonant cavity immersed in a static magnetic field.
The apparatus consists of a 136~$\ell$ cylindrical copper-plated stainless steel microwave cavity surrounded by a 7.5~T superconducting magnet.
The resonant frequency of the cavity is adjusted through the use of two movable internal bulk copper rods.
The RF power in the cavity is extracted using an antenna, and then amplified with a Josephson Parametric Amplifier (JPA)~\cite{Siddiqi_2004} and the following Heterostructure Field Effect Transistor (HFET) amplifiers~\cite{LNFLNC026A}.
For further details, see Ref.~\cite{doi:10.1063/5.0037857}.

In the present work, we will search for a \CaB signal in existing ADMX data that were collected with the explicit purpose of searching for axion dark matter.
The dataset amounts to a series of power measurements taken over 100-second intervals throughout the day, each taken at different resonant frequencies, and scanned in search of the unknown dark matter mass.
For each of these measurements, a dark matter signal would emerge as a narrow line in the data on top of a broad thermal photon background.
There is then a relatively clean separation into frequency regions where there is background only and those where there is signal and background, which allows one to calibrate the signal strength with respect to the thermal noise within one digitization bandwidth.
Such a separation is not possible for the \CaB.
The signal is broader than both the cavity linewidth and the digitization bandwidth: the only way to distinguish a \CaB signal from the thermal background in a single scan would be to exploit the $B^2$ scaling in Eq.~\eqref{eq:dPdw}.
If we combine multiple scans taken throughout the day, however, we can search for the daily modulation of the \CaB signal.
Unfortunately, the backgrounds also vary with time; there is time variation in the gain as well as temperature drifts in the system which are mentioned later.
The HFET response is relatively stable, about 2\% of fluctuation through a week, and therefore we did not apply any calibration, we rather add the fluctuation as systematic uncertainty as shown later.
The bias current of the JPA is adjusted every four digitizations to maximize the dark matter signal-to-noise ratio.
During this process, gain stability is not the figure of merit, and the gain can vary between 15--30~dB. 
This will lead to large fluctuations in the observed power as a function of time, and will form an important background when searching for a modulating signal.

\begin{figure}[!t]
\centering
\includegraphics[width=\linewidth]{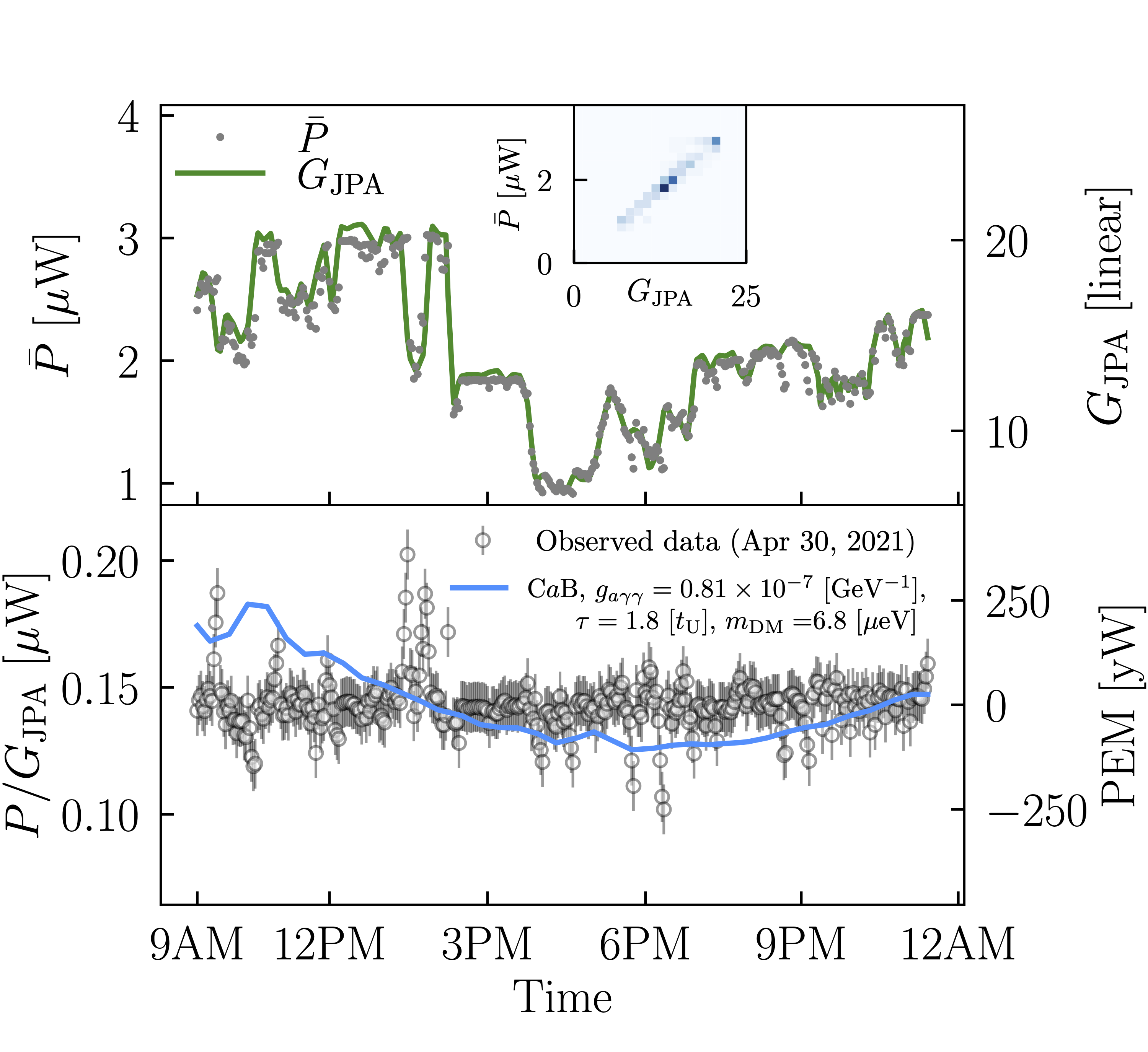}
\vspace{-0.65cm}
\caption{{\it Top:} an example of the mean power (gray) as a function of time collected on April 30, 2021.
We further show the JPA gain in green, and in the inset demonstrate the two are strongly correlated.
{\it Bottom:} For the same dataset, we divide the mean power out by the gain (left axis) and show the resulting PEM computed according to Eq.~\eqref{eq:powerexess} (right axis, gray circle). The light blue line shows CaB PEM as a reference. Parameters are chosen to be visible in this scale. A peak of $P/G_{JPA}$ at around 1-2 PM is due to the gain instability of the JPA caused by the liquid helium refill of the ADMX reservoir.   
}
\vspace{-0.5cm}
\label{fig:modulationEx}
\end{figure}

To mitigate the impact of gain fluctuations, we define a new observable, which we refer to as the Power Excess Modulation (PEM),
\begin{equation}
\hspace{-0.15cm}{\rm PEM} = \frac{\overline{P}(t)/G_{\rm JPA}(t)-\langle\overline{P}(t)/G_{\rm JPA}(t)\rangle}{\langle \sigma_{P}(t)/G_{\rm JPA}(t) \rangle}k_{\rm B}T_{\rm sys}\sqrt{\frac{b}{T}}.
\label{eq:powerexess}
\end{equation}
Here, $\overline{P}(t)$ and $\sigma_P(t)$ are the mean and standard deviation of the power in a dataset collected at a time $t$, computed over the frequencies in the digitization bandwidth, which is 100 Hz for the data we used.
For that same dataset, we further use the gain in the JPA, denoted by $G_{\rm JPA}(t)$.
Finally, $\langle \cdot \rangle$ corresponds to expressions averaged over a full day, \Tsys is the system-noise temperature, typically around 500~mK, \kboltz is the Boltzmann constant, $b$ is the bin width of the spectrum, and $T$ is the collection time for the dataset, which for the results we use is 100 seconds.

The PEM in Eq.~\eqref{eq:powerexess} quantifies the variation of the power in a single scan relative to the average for that day, accounting for the measured gain.
If we have PEM values collected at a series of times $t_i$, we can combine these into a single dataset $d_{\rm PEM} \equiv \{{\rm PEM} (t_i)\}$, within which we can look for an axion signal oscillating throughout the day.
To provide an example of the data, in Fig.~\ref{fig:modulationEx} (top) we show the mean power as a function of time throughout the day, for data taken on April 30, 2021.
As can be seen, the gain from the JPA varies by an ${\cal O}(1)$ value throughout the day, and those fluctuations are strongly correlated with the observed value of $\bar{P}$.
In the bottom panel, we show the power after the gain has been divided out, and it is within this data we search for a signal of the \CaB, which is also shown.
We note that the use of $\overline{P}(t)/G_{\rm JPA}(t)$ rather than the average power greatly improves the sensitivity to the \CaB.
Nevertheless, our knowledge of $G_{\rm JPA}(t)$ is imperfect.
Since measurements of the JPA gain were only performed approximately every 5 minutes, meaning values are not available for every dataset.
Therefore, residual variations caused by the JPA were taken into account as a systematic uncertainty described later.

\vspace{0.2cm}
\noindent {\bf Data reduction and systematics.}
%
We now describe the existing ADMX data we will use to search for the \CaB.
In particular, we made use of an existing ADMX search for dark matter performed over the frequency range 800--1020~MHz~\cite{PhysRevLett.127.261803}.
Given the background considerations are different when searching for the \CaB rather than dark matter, we applied additional quality cuts to the dataset.
Firstly, the JPA gain was found to be significantly unstable during the time when the data in the 995--1020~MHz frequency range were collected, and consequently we decided to exclude this entirely.
Secondly, we discarded all rescan datasets -- measurements performed to follow up any potential candidates observed in the dark matter search -- to avoid discontinuities in the PEM.
Even though these could be accounted for, the amount of rescan data is negligible, so including it would not substantially improve our sensitivity.
Finally, we discarded those datasets which were collected while the liquid helium was being refilled, as the vibrations led to further fluctuations in the JPA gain.
These three cuts reduced our initial dataset by roughly 50\%.

Before analyzing the data, there are three sources of systematic uncertainty we identified that are beyond those conventionally accounted for in haloscope analyses \footnote{Representative uncertainties are related to electric field distribution, noise temperature measurements, and cavity parameters, see Ref. \cite{PhysRevLett.127.261803}.}, as they are unique to the PEM analysis: short timescale fluctuations dominated by the JPA variation, power fluctuations from the other parts of the radio frequency (RF) system, and an imperfect calculation of $K(\omega,\alpha)$.
We describe each of these below in turn.

To begin with, as mentioned the $\bar{P}(t)$ values we measure are collected every $\sim$100 seconds, whereas JPA gain variation are only remeasured every $\sim$5 minutes.
The two are strongly correlated, however, as clearly shown in Fig.~\ref{fig:modulationEx}, and therefore the imperfect knowledge of $G_{\rm JPA}$ represents an uncertainty to be accounted for.
The gain fluctuations are statistically random and approximately Gaussian.
In order to model the systematic error induced by these fluctuations, we take the full PEM dataset as a function of time and fit it with a Savitzky-Golay (SG) filter (length 201 and polynomial order 3).
The SG filter is used to remove all fluctuations over timescales larger than $\sim$5 minutes, for instance, a potential HFET gain drift or even a possible contribution from the \CaB.
The remaining short-time fluctuations are attributed to the JPA, and we model these as a Gaussian with a width determined by the standard deviation of these residuals.
This fluctuation is then combined with the other systematic uncertainties we account for in quadrature.
We emphasize that the SG filter is used only for estimating the uncertainty; the fit to search for a \CaB amplitude is performed without the SG filter.

While variations in the JPA gain are the dominant source of temporal variations in the power, there are other milder contributions from the RF system, such as variations in temperature, although the leading second-order effect is gain fluctuations in the HFET.
This uncertainty was estimated by directly measuring the stability of the room-temperature RF system.
During the measurements, the part of the RF system that is cooled down was disconnected from the full apparatus, so that the fluctuations were only measured in the latter.
Through this process, the fluctuations were measured to be at most 2\%, and this value was incorporated as a systematic uncertainty.
There is, however, an additional HFET amplifier in the part of the instrument that is cooled. Therefore, we disconnected it from the above measurement, and we conservatively assign the same 2\% uncertainty to the additional HFET amplifier.
Combining the two in quadrature, we then attribute the full RF system (excluding the JPA) a systematic uncertainty of 2.8\%.

The final source of systematic uncertainty arises from the calculation of $K(\omega,\alpha)$.
We determined this from Eq.~\eqref{eq:K} though that result assumes a perfectly cylindrical cavity.
For comparison, the relevant form factor for dark matter, $C$, deviates from this perfect value due to the presence of the tuning rods in the cavity, for example, and is precisely determined using simulations with Ansys HFSS.
We attribute an uncertainty of 30\% to our computation of $K$, as this is the magnitude of the variation in $C$ as the tuning rods are moved confirmed by simulation with Ansys HFSS.

\vspace{0.2cm}
\noindent {\bf A C\textit{a}B search in existing ADMX data.}
%
Having described our dataset, \dPEM, and the PEM analysis framework, we turn to the search for the \CaB.
We will perform the search using a Gaussian likelihood analysis, determining limits with a likelihood ratio test statistic.
While we expect fluctuations in the power throughout the day from various background sources as described above, these have been accounted for either by our explicit inclusion of $G_{\rm JPA}$ in the PEM or else through the use of systematic uncertainties.
Accordingly, our null hypothesis ($H_0$) is simply that ${\rm PEM}=0$.
For our alternative hypothesis ($H_1$) we add to the null hypothesis a contribution from the \CaB.
As a reference point, we consider the largest \CaB signal that can be presently obtained: this occurs when we take $\tau = 1.8\,\tU$, and $\gaCAST = 0.66 \times 10^{-10}~{\rm GeV}^{-1}$, i.e. saturating the DES and CAST 95\% confidence limits, respectively (with the latter taken from Ref.~\cite{Anastassopoulos2017}).
From this maximally allowed hypothesis we can then vary $\ga$, which varies the overall amplitude of the signal according to Eq.~\eqref{eq:dPdw}, determine the best fit value and then define a signal strength $\mu = (\ga^{\rm fit}/\gaCAST)^2$, which will also be the ratio of the best fit PEM to that predicted when $\mu=1$.
Note we choose to vary $\ga$, rather than $\tau$, as the former solely scales the signal amplitude; the lifetime changes the shape of the signal as we describe in the supplementary material.
From here we define a signal hypothesis $H_1$ as the alternative hypothesis where $\mu$ obtains its preferred value in the data.

From these hypotheses, we define the probability of $H_0$ and $H_1$ as $p(H_0)$ and $p(H_1)$, respectively.
We then set the following criteria:
\begin{itemize}
    \item $p(H_0)>0.003$: no evidence for new physics;
    \item $p(H_0)<0.003$ and $p(H_1)<0.003$: the null hypothesis is disfavored, but there is no evidence for the \CaB; and
    \item $p(H_0)<0.003$ and $p(H_1)>0.003$: there is a preference for the \CaB.
\end{itemize}
In the above, 0.003 corresponds to the 3$\sigma$ threshold.
In the absence of evidence for the maximal \CaB signal predicted by $H_1$, we can determine a 95\% confidence level (C.L.) limit on $\ga$ by scaling $\gaCAST$ with $\sqrt{\mu + 1.644\times \Delta\mu}$, where $\Delta\mu$ is fit uncertainty on the best fit value of $\mu$.
The value of 1.644 arises from 95\% of the one-sided Gaussian probability.

\begin{figure}[!t]
\centering
\includegraphics[width=\linewidth]{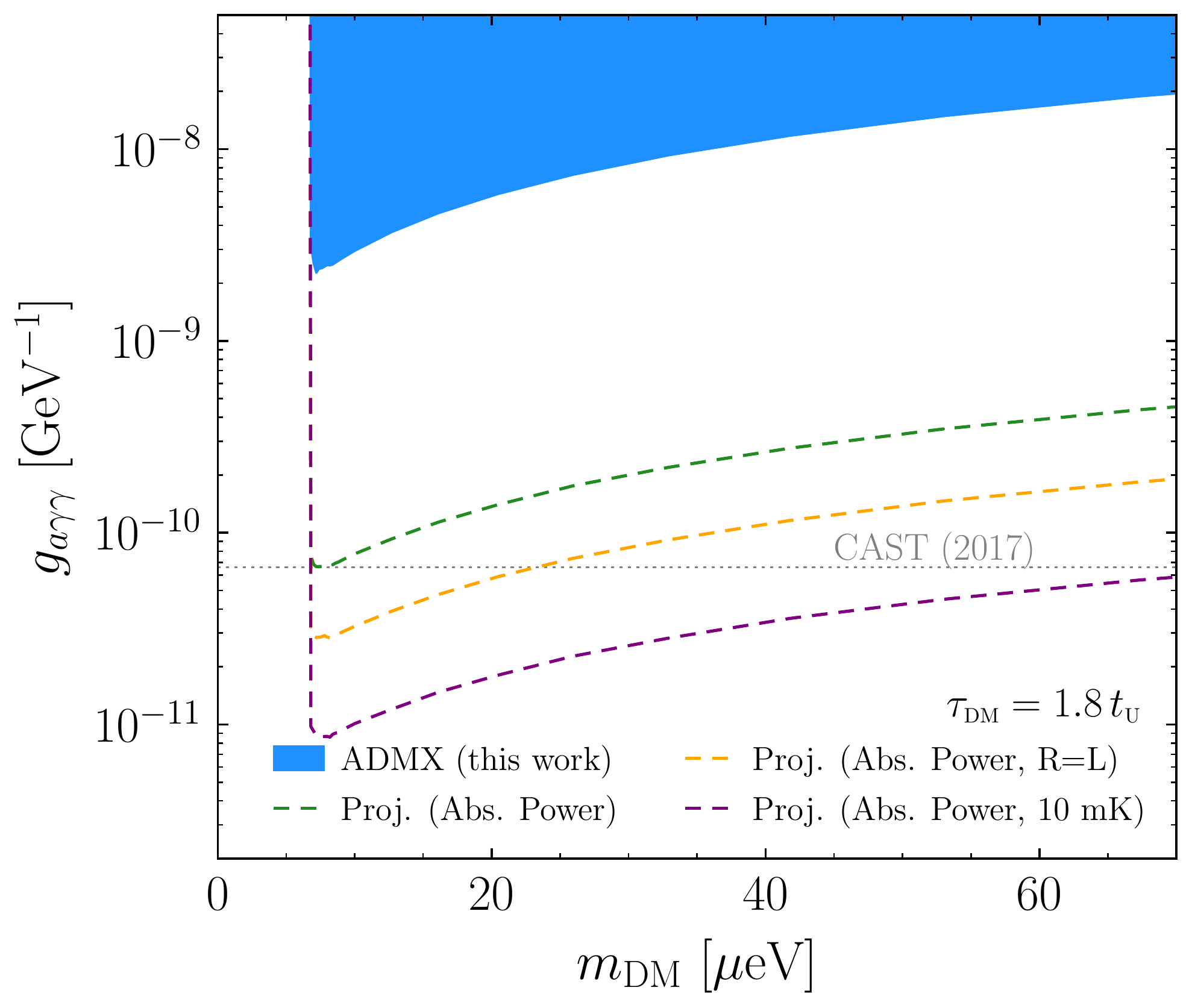}
\vspace{-0.6cm}
\caption{95\% C.L. on $\ga$ for a \CaB arising from the decay of dark matter, assuming $\tau=1.8\,\tU$.
The horizontal dashed line shows the CAST bound~\cite{Anastassopoulos2017} which assumes general axion-photon coupling from axion production in the Sun. The results do not yet exceed the CAST bound, and are primarily limited by fluctuations in the JPA gain.
These would be removed if the same analysis were performed using an absolute power measurement, and with other improvements such as a reduced system temperature or modified cavity geometry, significantly enhanced sensitivity can be achieved.
See text for details.
}
\vspace{-0.5cm}
\label{fig:limits}
\end{figure}

Using the above procedure, we searched for a signal of the \CaB between 800 and 995~MHz, which in total amounted to 143 days of data.
Nevertheless, we found that the data taken near 850~MHz dominated the sensitivity as it was taken over a period of time where the JPA was especially stable.
We analyzed each day of data separately and found that all spectra were consistent with the expected background in the absence of a \CaB, i.e. $p(H_0)>0.003$. 
Therefore, we use the combined data over all days to establish an upper limit on $\ga$.
The results are shown in Fig.~\ref{fig:limits} with $\tau=1.8\,\tU$; results for larger values of $\tau$, for which our constraints on $g_{a\gamma\gamma}$ degrade, are provided in the supplementary material.

At present, this first search for the \CaB is not able to reach the allowed parameter space below $\gaCAST$.
The primary limitation is that existing haloscopes are only sensitive to the relative power excess, for instance, as encapsulated in the PEM.
However, this is not a fundamental limitation.
For example, a single photon counter using a superconducting qubit (see e.g. Ref.~\cite{PhysRevLett.126.141302}) essentially measures the absolute photon occupation number in the cavity, which would allow for absolute power measurements. In this case, thermal excitation or false positive rate of qubits dictates the background level.
Such a measurement would remove the need to subtract the mean, as we do in the PEM given we are presently only sensitive to time variations rather than the absolute scale.
As a result, at present we are completely insensitive to the contribution to the \CaB from extragalactic decays, worsening the sensitivity to $g_{a\gamma\gamma}$ by $\sim$$2/5$.
An additional benefit of an absolute power measurement is that as it only measures the number of states in the cavity it will render the amplifiers gain fluctuations negligible when at present they are our dominant background.
Removing this background would improve our sensitivity to roughly $\gaCAST$.
In addition to measuring the absolute power, single photon counters using a superconducting qubit can potentially achieve a noise temperature of around 10 mK by using quantum nondemolution measurements~\cite{PhysRevLett.65.976}, which would improve the sensitivity further.
Finally, as we describe in the supplementary material, the ADMX cavity geometry ($L \sim 5 R$) significantly suppresses the \CaB power through $K(\omega_0,\alpha)$.
A cavity with $L \sim R$ would largely lift that suppression.
In Fig.~\ref{fig:limits} we show the expected improved sensitivity that would result from each of these considerations.

\vspace{0.2cm}
\noindent {\bf Discussion.}
%
We have performed the first direct search for the \CaB with an axion haloscope.
While there are many forms the \CaB could take, we focused on the possibility of a cascade decay of dark matter to axions, $\chi \to \varphi \varphi \to aaaa$, and exploited the resulting daily modulation in the signal.
In particular, we introduced the PEM analysis to search for a daily variation in the power, and then applied this method to existing ADMX data.

While our results are specific to the dark matter decay \CaB, the methodology employed here is general and can be used to search for other possible broadband signals in axion haloscopes.
The present sensitivity we obtained with the PEM is roughly an order of magnitude weaker than might have been expected, and this was solely due to the large variations in the gain.
To improve on this, future measurements can monitor the relative HFET gain fluctuation, for instance, by injecting RF tones during data taking and thereby have an accurate measurement of the gain at all times rather than every 5 minutes.
Also, data-taking without fine-tuning of the JPA bias current at the same cavity resonant frequency would realize stabler PEM spectra. 
Separately, looking even further forward, the sensitivity will improve significantly once absolute power measurements in the cavity are possible, an avenue that will be opened by single photon measurements.
Combining this improvement with the expected strides axion haloscopes will make in the coming decade, it is likely this will be only the first of many searches for non-dark matter signals with such instruments.
Indeed, it could well be that the first signs of new physics in these instruments emerges in the form of a \CaB, high-frequency gravitational waves (see e.g. Refs.~\cite{berlin2021detecting,Domcke:2022rgu}), or another as yet unanticipated signal.

\vspace{0.5 cm}
\noindent {\it Acknowledgements.}
The work presented benefited from discussions with members of the HAYSTAC Collaboration, in particular Maryam H. Esmat, Sumita Ghosh, Alexander F. Leder, and Danielle Speller.
This work was supported by the U.S. Department of Energy through Grants No. DE-SC0009800, No. DESC0009723, No. DE-SC0010296, No. DE-SC0010280, No. DE-SC0011665, No. DEFG02-97ER41029, No. DEFG02-96ER40956, No. DEC03-76SF00098, and No. DE-SC0017987.
Fermilab is a U.S. Department of Energy, Office of Science, HEP User Facility. Fermilab is managed by Fermi Research Alliance, LLC (FRA), acting under Contract No. DE-AC02-07CH11359.
Additional support was provided by the Heising-Simons Foundation and by the Lawrence Livermore National Laboratory LDRD office. Pacific Northwest National Laboratory is a multi-program national laboratory operated for the U.S. DOE by Battelle Memorial Institute under Contract No. DE-AC05-76RL01830. Prepared in part by LLNL under Contract DE-AC52-07NA27344, Release No. LLNL-JRNL-845987”.
The corresponding author is supported by JSPS Overseas Research Fellowships No. 202060305.
Chelsea Bartram acknowledges support from the Panofsky Fellowship at SLAC.

\bibliographystyle{apsrev4-2}
\bibliography{references}

\providecommand{\noopsort}[1]{}\providecommand{\singleletter}[1]{#1}%
\begin{thebibliography}{54}%
\makeatletter
\providecommand \@ifxundefined [1]{%
 \@ifx{#1\undefined}
}%
\providecommand \@ifnum [1]{%
 \ifnum #1\expandafter \@firstoftwo
 \else \expandafter \@secondoftwo
 \fi
}%
\providecommand \@ifx [1]{%
 \ifx #1\expandafter \@firstoftwo
 \else \expandafter \@secondoftwo
 \fi
}%
\providecommand \natexlab [1]{#1}%
\providecommand \enquote  [1]{``#1''}%
\providecommand \bibnamefont  [1]{#1}%
\providecommand \bibfnamefont [1]{#1}%
\providecommand \citenamefont [1]{#1}%
\providecommand \href@noop [0]{\@secondoftwo}%
\providecommand \href [0]{\begingroup \@sanitize@url \@href}%
\providecommand \@href[1]{\@@startlink{#1}\@@href}%
\providecommand \@@href[1]{\endgroup#1\@@endlink}%
\providecommand \@sanitize@url [0]{\catcode `\\12\catcode `\$12\catcode
  `\&12\catcode `\#12\catcode `\^12\catcode `\_12\catcode `\%12\relax}%
\providecommand \@@startlink[1]{}%
\providecommand \@@endlink[0]{}%
\providecommand \url  [0]{\begingroup\@sanitize@url \@url }%
\providecommand \@url [1]{\endgroup\@href {#1}{\urlprefix }}%
\providecommand \urlprefix  [0]{URL }%
\providecommand \Eprint [0]{\href }%
\providecommand \doibase [0]{https://doi.org/}%
\providecommand \selectlanguage [0]{\@gobble}%
\providecommand \bibinfo  [0]{\@secondoftwo}%
\providecommand \bibfield  [0]{\@secondoftwo}%
\providecommand \translation [1]{[#1]}%
\providecommand \BibitemOpen [0]{}%
\providecommand \bibitemStop [0]{}%
\providecommand \bibitemNoStop [0]{.\EOS\space}%
\providecommand \EOS [0]{\spacefactor3000\relax}%
\providecommand \BibitemShut  [1]{\csname bibitem#1\endcsname}%
\let\auto@bib@innerbib\@empty
\bibitem [{\citenamefont {Peccei}\ and\ \citenamefont
  {Quinn}(1977{\natexlab{a}})}]{PhysRevD.16.1791}%
  \BibitemOpen
  \bibfield  {author} {\bibinfo {author} {\bibfnamefont {R.~D.}\ \bibnamefont
  {Peccei}}\ and\ \bibinfo {author} {\bibfnamefont {H.~R.}\ \bibnamefont
  {Quinn}},\ }\href {https://doi.org/10.1103/PhysRevD.16.1791} {\bibfield
  {journal} {\bibinfo  {journal} {Phys. Rev. D}\ }\textbf {\bibinfo {volume}
  {16}},\ \bibinfo {pages} {1791} (\bibinfo {year}
  {1977}{\natexlab{a}})}\BibitemShut {NoStop}%
\bibitem [{\citenamefont {Weinberg}(1978)}]{PhysRevLett.40.223}%
  \BibitemOpen
  \bibfield  {author} {\bibinfo {author} {\bibfnamefont {S.}~\bibnamefont
  {Weinberg}},\ }\href {https://doi.org/10.1103/PhysRevLett.40.223} {\bibfield
  {journal} {\bibinfo  {journal} {Phys. Rev. Lett.}\ }\textbf {\bibinfo
  {volume} {40}},\ \bibinfo {pages} {223} (\bibinfo {year} {1978})}\BibitemShut
  {NoStop}%
\bibitem [{\citenamefont {Peccei}\ and\ \citenamefont
  {Quinn}(1977{\natexlab{b}})}]{Peccei1977June}%
  \BibitemOpen
  \bibfield  {author} {\bibinfo {author} {\bibfnamefont {R.~D.}\ \bibnamefont
  {Peccei}}\ and\ \bibinfo {author} {\bibfnamefont {H.~R.}\ \bibnamefont
  {Quinn}},\ }\href {https://doi.org/10.1103/PhysRevLett.38.1440} {\bibfield
  {journal} {\bibinfo  {journal} {Phys. Rev. Lett.}\ }\textbf {\bibinfo
  {volume} {38}},\ \bibinfo {pages} {1440} (\bibinfo {year}
  {1977}{\natexlab{b}})}\BibitemShut {NoStop}%
\bibitem [{\citenamefont {Wilczek}(1978)}]{PhysRevLett.40.279}%
  \BibitemOpen
  \bibfield  {author} {\bibinfo {author} {\bibfnamefont {F.}~\bibnamefont
  {Wilczek}},\ }\href {https://doi.org/10.1103/PhysRevLett.40.279} {\bibfield
  {journal} {\bibinfo  {journal} {Phys. Rev. Lett.}\ }\textbf {\bibinfo
  {volume} {40}},\ \bibinfo {pages} {279} (\bibinfo {year} {1978})}\BibitemShut
  {NoStop}%
\bibitem [{\citenamefont {Sikivie}(2021)}]{RevModPhys.93.015004}%
  \BibitemOpen
  \bibfield  {author} {\bibinfo {author} {\bibfnamefont {P.}~\bibnamefont
  {Sikivie}},\ }\href {https://doi.org/10.1103/RevModPhys.93.015004} {\bibfield
   {journal} {\bibinfo  {journal} {Rev. Mod. Phys.}\ }\textbf {\bibinfo
  {volume} {93}},\ \bibinfo {pages} {015004} (\bibinfo {year}
  {2021})}\BibitemShut {NoStop}%
\bibitem [{\citenamefont {Hook}(2019)}]{Hook:2018dlk}%
  \BibitemOpen
  \bibfield  {author} {\bibinfo {author} {\bibfnamefont {A.}~\bibnamefont
  {Hook}},\ }\href@noop {} {\bibfield  {journal} {\bibinfo  {journal} {PoS}\
  }\textbf {\bibinfo {volume} {TASI2018}},\ \bibinfo {pages} {004} (\bibinfo
  {year} {2019})},\ \Eprint {https://arxiv.org/abs/1812.02669}
  {arXiv:1812.02669 [hep-ph]} \BibitemShut {NoStop}%
\bibitem [{\citenamefont {Adams}\ \emph {et~al.}(2023)\citenamefont {Adams},
  \citenamefont {Aggarwal}, \citenamefont {Agrawal}, \citenamefont
  {Balafendiev}, \citenamefont {Bartram}, \citenamefont {Baryakhtar} \emph
  {et~al.}}]{adams2023axion}%
  \BibitemOpen
  \bibfield  {author} {\bibinfo {author} {\bibfnamefont {C.~B.}\ \bibnamefont
  {Adams}}, \bibinfo {author} {\bibfnamefont {N.}~\bibnamefont {Aggarwal}},
  \bibinfo {author} {\bibfnamefont {A.}~\bibnamefont {Agrawal}}, \bibinfo
  {author} {\bibfnamefont {R.}~\bibnamefont {Balafendiev}}, \bibinfo {author}
  {\bibfnamefont {C.}~\bibnamefont {Bartram}}, \bibinfo {author} {\bibfnamefont
  {M.}~\bibnamefont {Baryakhtar}}, \emph {et~al.},\ }\href@noop {} {\bibinfo
  {title} {Axion dark matter}} (\bibinfo {year} {2023}),\ \Eprint
  {https://arxiv.org/abs/2203.14923} {arXiv:2203.14923 [hep-ex]} \BibitemShut
  {NoStop}%
\bibitem [{\citenamefont {Dror}\ \emph {et~al.}(2021)\citenamefont {Dror},
  \citenamefont {Murayama},\ and\ \citenamefont {Rodd}}]{PhysRevD.103.115004}%
  \BibitemOpen
  \bibfield  {author} {\bibinfo {author} {\bibfnamefont {J.~A.}\ \bibnamefont
  {Dror}}, \bibinfo {author} {\bibfnamefont {H.}~\bibnamefont {Murayama}},\
  and\ \bibinfo {author} {\bibfnamefont {N.~L.}\ \bibnamefont {Rodd}},\ }\href
  {https://doi.org/10.1103/PhysRevD.103.115004} {\bibfield  {journal} {\bibinfo
   {journal} {Phys. Rev. D}\ }\textbf {\bibinfo {volume} {103}},\ \bibinfo
  {pages} {115004} (\bibinfo {year} {2021})}\BibitemShut {NoStop}%
\bibitem [{\citenamefont {Abazajian}\ \emph {et~al.}(2016)\citenamefont
  {Abazajian} \emph {et~al.}}]{CMB-S4:2016ple}%
  \BibitemOpen
  \bibfield  {author} {\bibinfo {author} {\bibfnamefont {K.~N.}\ \bibnamefont
  {Abazajian}} \emph {et~al.} (\bibinfo {collaboration} {CMB-S4}),\ }\href@noop
  {} {\  (\bibinfo {year} {2016})},\ \Eprint {https://arxiv.org/abs/1610.02743}
  {arXiv:1610.02743 [astro-ph.CO]} \BibitemShut {NoStop}%
\bibitem [{\citenamefont {Baumann}\ \emph {et~al.}(2016)\citenamefont
  {Baumann}, \citenamefont {Green},\ and\ \citenamefont
  {Wallisch}}]{PhysRevLett.117.171301}%
  \BibitemOpen
  \bibfield  {author} {\bibinfo {author} {\bibfnamefont {D.}~\bibnamefont
  {Baumann}}, \bibinfo {author} {\bibfnamefont {D.}~\bibnamefont {Green}},\
  and\ \bibinfo {author} {\bibfnamefont {B.}~\bibnamefont {Wallisch}},\ }\href
  {https://doi.org/10.1103/PhysRevLett.117.171301} {\bibfield  {journal}
  {\bibinfo  {journal} {Phys. Rev. Lett.}\ }\textbf {\bibinfo {volume} {117}},\
  \bibinfo {pages} {171301} (\bibinfo {year} {2016})}\BibitemShut {NoStop}%
\bibitem [{\citenamefont {Verde}\ \emph {et~al.}(2019)\citenamefont {Verde},
  \citenamefont {Treu},\ and\ \citenamefont {Riess}}]{Verde2019}%
  \BibitemOpen
  \bibfield  {author} {\bibinfo {author} {\bibfnamefont {L.}~\bibnamefont
  {Verde}}, \bibinfo {author} {\bibfnamefont {T.}~\bibnamefont {Treu}},\ and\
  \bibinfo {author} {\bibfnamefont {A.~G.}\ \bibnamefont {Riess}},\ }\href
  {https://doi.org/10.1038/s41550-019-0902-0} {\bibfield  {journal} {\bibinfo
  {journal} {Nature Astronomy}\ }\textbf {\bibinfo {volume} {3}},\ \bibinfo
  {pages} {891} (\bibinfo {year} {2019})}\BibitemShut {NoStop}%
\bibitem [{\citenamefont {Sikivie}(1983)}]{PhysRevLett.51.1415}%
  \BibitemOpen
  \bibfield  {author} {\bibinfo {author} {\bibfnamefont {P.}~\bibnamefont
  {Sikivie}},\ }\href {https://doi.org/10.1103/PhysRevLett.51.1415} {\bibfield
  {journal} {\bibinfo  {journal} {Phys. Rev. Lett.}\ }\textbf {\bibinfo
  {volume} {51}},\ \bibinfo {pages} {1415} (\bibinfo {year}
  {1983})}\BibitemShut {NoStop}%
\bibitem [{\citenamefont {Ouellet}\ \emph {et~al.}(2019)\citenamefont {Ouellet}
  \emph {et~al.}}]{PhysRevD.99.052012}%
  \BibitemOpen
  \bibfield  {author} {\bibinfo {author} {\bibfnamefont {J.~L.}\ \bibnamefont
  {Ouellet}} \emph {et~al.},\ }\href
  {https://doi.org/10.1103/PhysRevD.99.052012} {\bibfield  {journal} {\bibinfo
  {journal} {Phys. Rev. D}\ }\textbf {\bibinfo {volume} {99}},\ \bibinfo
  {pages} {052012} (\bibinfo {year} {2019})}\BibitemShut {NoStop}%
\bibitem [{\citenamefont {Gelmini}\ \emph {et~al.}(2020)\citenamefont
  {Gelmini}, \citenamefont {Millar}, \citenamefont {Takhistov},\ and\
  \citenamefont {Vitagliano}}]{Gelmini:2020kcu}%
  \BibitemOpen
  \bibfield  {author} {\bibinfo {author} {\bibfnamefont {G.~B.}\ \bibnamefont
  {Gelmini}}, \bibinfo {author} {\bibfnamefont {A.~J.}\ \bibnamefont {Millar}},
  \bibinfo {author} {\bibfnamefont {V.}~\bibnamefont {Takhistov}},\ and\
  \bibinfo {author} {\bibfnamefont {E.}~\bibnamefont {Vitagliano}},\ }\href
  {https://doi.org/10.1103/PhysRevD.102.043003} {\bibfield  {journal} {\bibinfo
   {journal} {Phys. Rev. D}\ }\textbf {\bibinfo {volume} {102}},\ \bibinfo
  {pages} {043003} (\bibinfo {year} {2020})},\ \Eprint
  {https://arxiv.org/abs/2006.06836} {arXiv:2006.06836 [hep-ph]} \BibitemShut
  {NoStop}%
\bibitem [{\citenamefont {Arvanitaki}\ and\ \citenamefont
  {Geraci}(2014)}]{arvanitaki2014resonantly}%
  \BibitemOpen
  \bibfield  {author} {\bibinfo {author} {\bibfnamefont {A.}~\bibnamefont
  {Arvanitaki}}\ and\ \bibinfo {author} {\bibfnamefont {A.~A.}\ \bibnamefont
  {Geraci}},\ }\href@noop {} {\bibfield  {journal} {\bibinfo  {journal}
  {Physical review letters}\ }\textbf {\bibinfo {volume} {113}},\ \bibinfo
  {pages} {161801} (\bibinfo {year} {2014})}\BibitemShut {NoStop}%
\bibitem [{\citenamefont {Liu}\ \emph {et~al.}(2022)\citenamefont {Liu},
  \citenamefont {Dona}, \citenamefont {Hoshino}, \citenamefont {Knirck},
  \citenamefont {Kurinsky}, \citenamefont {Malaker} \emph
  {et~al.}}]{BREAD:2021tpx}%
  \BibitemOpen
  \bibfield  {author} {\bibinfo {author} {\bibfnamefont {J.}~\bibnamefont
  {Liu}}, \bibinfo {author} {\bibfnamefont {K.}~\bibnamefont {Dona}}, \bibinfo
  {author} {\bibfnamefont {G.}~\bibnamefont {Hoshino}}, \bibinfo {author}
  {\bibfnamefont {S.}~\bibnamefont {Knirck}}, \bibinfo {author} {\bibfnamefont
  {N.}~\bibnamefont {Kurinsky}}, \bibinfo {author} {\bibfnamefont
  {M.}~\bibnamefont {Malaker}}, \emph {et~al.} (\bibinfo {collaboration}
  {BREAD}),\ }\href {https://doi.org/10.1103/PhysRevLett.128.131801} {\bibfield
   {journal} {\bibinfo  {journal} {Phys. Rev. Lett.}\ }\textbf {\bibinfo
  {volume} {128}},\ \bibinfo {pages} {131801} (\bibinfo {year} {2022})},\
  \Eprint {https://arxiv.org/abs/2111.12103} {arXiv:2111.12103
  [physics.ins-det]} \BibitemShut {NoStop}%
\bibitem [{\citenamefont {Aybas}\ \emph {et~al.}(2021)\citenamefont {Aybas},
  \citenamefont {Adam}, \citenamefont {Blumenthal}, \citenamefont {Gramolin},
  \citenamefont {Johnson}, \citenamefont {Kleyheeg}, \citenamefont {Afach},\
  and\ \citenamefont {othors}}]{Aybas2021a}%
  \BibitemOpen
  \bibfield  {author} {\bibinfo {author} {\bibfnamefont {D.}~\bibnamefont
  {Aybas}}, \bibinfo {author} {\bibfnamefont {J.}~\bibnamefont {Adam}},
  \bibinfo {author} {\bibfnamefont {E.}~\bibnamefont {Blumenthal}}, \bibinfo
  {author} {\bibfnamefont {A.~V.}\ \bibnamefont {Gramolin}}, \bibinfo {author}
  {\bibfnamefont {D.}~\bibnamefont {Johnson}}, \bibinfo {author} {\bibfnamefont
  {A.}~\bibnamefont {Kleyheeg}}, \bibinfo {author} {\bibfnamefont
  {S.}~\bibnamefont {Afach}},\ and\ \bibinfo {author} {\bibnamefont {othors}},\
  }\href {https://doi.org/10.1103/PhysRevLett.126.141802} {\bibfield  {journal}
  {\bibinfo  {journal} {Phys. Rev. Lett.}\ }\textbf {\bibinfo {volume} {126}},\
  \bibinfo {pages} {141802} (\bibinfo {year} {2021})}\BibitemShut {NoStop}%
\bibitem [{\citenamefont {Wu}\ \emph {et~al.}(2019)\citenamefont {Wu},
  \citenamefont {Blanchard}, \citenamefont {Centers}, \citenamefont {Figueroa},
  \citenamefont {Garcon}, \citenamefont {Graham}, \citenamefont {Kimball},
  \citenamefont {Rajendran}, \citenamefont {Stadnik}, \citenamefont {Sushkov},
  \citenamefont {Wickenbrock},\ and\ \citenamefont {Budker}}]{Wu:2019AxionDM}%
  \BibitemOpen
  \bibfield  {author} {\bibinfo {author} {\bibfnamefont {T.}~\bibnamefont
  {Wu}}, \bibinfo {author} {\bibfnamefont {J.~W.}\ \bibnamefont {Blanchard}},
  \bibinfo {author} {\bibfnamefont {G.~P.}\ \bibnamefont {Centers}}, \bibinfo
  {author} {\bibfnamefont {N.~L.}\ \bibnamefont {Figueroa}}, \bibinfo {author}
  {\bibfnamefont {A.}~\bibnamefont {Garcon}}, \bibinfo {author} {\bibfnamefont
  {P.~W.}\ \bibnamefont {Graham}}, \bibinfo {author} {\bibfnamefont {D.~F.~J.}\
  \bibnamefont {Kimball}}, \bibinfo {author} {\bibfnamefont {S.}~\bibnamefont
  {Rajendran}}, \bibinfo {author} {\bibfnamefont {Y.~V.}\ \bibnamefont
  {Stadnik}}, \bibinfo {author} {\bibfnamefont {A.~O.}\ \bibnamefont
  {Sushkov}}, \bibinfo {author} {\bibfnamefont {A.}~\bibnamefont
  {Wickenbrock}},\ and\ \bibinfo {author} {\bibfnamefont {D.}~\bibnamefont
  {Budker}},\ }\href {https://doi.org/10.1103/PhysRevLett.122.191302}
  {\bibfield  {journal} {\bibinfo  {journal} {Phys. Rev. Lett.}\ }\textbf
  {\bibinfo {volume} {122}},\ \bibinfo {pages} {191302} (\bibinfo {year}
  {2019})}\BibitemShut {NoStop}%
\bibitem [{\citenamefont {Brouwer}\ \emph {et~al.}(2022)\citenamefont {Brouwer}
  \emph {et~al.}}]{Brouwer:2022bwo}%
  \BibitemOpen
  \bibfield  {author} {\bibinfo {author} {\bibfnamefont {L.}~\bibnamefont
  {Brouwer}} \emph {et~al.},\ }\href@noop {} {\  (\bibinfo {year} {2022})},\
  \Eprint {https://arxiv.org/abs/2203.11246} {arXiv:2203.11246 [hep-ex]}
  \BibitemShut {NoStop}%
\bibitem [{\citenamefont {Salemi}\ \emph {et~al.}(2021)\citenamefont {Salemi}
  \emph {et~al.}}]{Salemi:2021gck}%
  \BibitemOpen
  \bibfield  {author} {\bibinfo {author} {\bibfnamefont {C.~P.}\ \bibnamefont
  {Salemi}} \emph {et~al.},\ }\href
  {https://doi.org/10.1103/PhysRevLett.127.081801} {\bibfield  {journal}
  {\bibinfo  {journal} {Phys. Rev. Lett.}\ }\textbf {\bibinfo {volume} {127}},\
  \bibinfo {pages} {081801} (\bibinfo {year} {2021})},\ \Eprint
  {https://arxiv.org/abs/2102.06722} {arXiv:2102.06722 [hep-ex]} \BibitemShut
  {NoStop}%
\bibitem [{\citenamefont {Backes}\ \emph {et~al.}(2021)\citenamefont {Backes}
  \emph {et~al.}}]{Backes2021}%
  \BibitemOpen
  \bibfield  {author} {\bibinfo {author} {\bibfnamefont {K.~M.}\ \bibnamefont
  {Backes}} \emph {et~al.},\ }\href
  {https://doi.org/10.1038/s41586-021-03226-7} {\bibfield  {journal} {\bibinfo
  {journal} {Nature}\ }\textbf {\bibinfo {volume} {590}},\ \bibinfo {pages}
  {238} (\bibinfo {year} {2021})}\BibitemShut {NoStop}%
\bibitem [{\citenamefont {Baryakhtar}\ \emph {et~al.}(2018)\citenamefont
  {Baryakhtar}, \citenamefont {Huang},\ and\ \citenamefont
  {Lasenby}}]{Baryakhtar:2018doz}%
  \BibitemOpen
  \bibfield  {author} {\bibinfo {author} {\bibfnamefont {M.}~\bibnamefont
  {Baryakhtar}}, \bibinfo {author} {\bibfnamefont {J.}~\bibnamefont {Huang}},\
  and\ \bibinfo {author} {\bibfnamefont {R.}~\bibnamefont {Lasenby}},\ }\href
  {https://doi.org/10.1103/PhysRevD.98.035006} {\bibfield  {journal} {\bibinfo
  {journal} {Phys. Rev. D}\ }\textbf {\bibinfo {volume} {98}},\ \bibinfo
  {pages} {035006} (\bibinfo {year} {2018})},\ \Eprint
  {https://arxiv.org/abs/1803.11455} {arXiv:1803.11455 [hep-ph]} \BibitemShut
  {NoStop}%
\bibitem [{\citenamefont {Brun}\ \emph {et~al.}(2019)\citenamefont {Brun} \emph
  {et~al.}}]{Brun:2019lyf}%
  \BibitemOpen
  \bibfield  {author} {\bibinfo {author} {\bibfnamefont {P.}~\bibnamefont
  {Brun}} \emph {et~al.} (\bibinfo {collaboration} {MADMAX}),\ }\href
  {https://doi.org/10.1140/epjc/s10052-019-6683-x} {\bibfield  {journal}
  {\bibinfo  {journal} {Eur. Phys. J.}\ }\textbf {\bibinfo {volume} {C79}},\
  \bibinfo {pages} {186} (\bibinfo {year} {2019})},\ \Eprint
  {https://arxiv.org/abs/1901.07401} {arXiv:1901.07401 [physics.ins-det]}
  \BibitemShut {NoStop}%
\bibitem [{\citenamefont {Gramolin}\ \emph {et~al.}(2021)\citenamefont
  {Gramolin}, \citenamefont {Aybas}, \citenamefont {Johnson}, \citenamefont
  {Adam},\ and\ \citenamefont {Sushkov}}]{Gramolin2021}%
  \BibitemOpen
  \bibfield  {author} {\bibinfo {author} {\bibfnamefont {A.~V.}\ \bibnamefont
  {Gramolin}}, \bibinfo {author} {\bibfnamefont {D.}~\bibnamefont {Aybas}},
  \bibinfo {author} {\bibfnamefont {D.}~\bibnamefont {Johnson}}, \bibinfo
  {author} {\bibfnamefont {J.}~\bibnamefont {Adam}},\ and\ \bibinfo {author}
  {\bibfnamefont {A.~O.}\ \bibnamefont {Sushkov}},\ }\href
  {https://doi.org/10.1038/s41567-020-1006-6} {\bibfield  {journal} {\bibinfo
  {journal} {Nature Physics}\ }\textbf {\bibinfo {volume} {17}},\ \bibinfo
  {pages} {79} (\bibinfo {year} {2021})}\BibitemShut {NoStop}%
\bibitem [{\citenamefont {Irastorza}\ \emph {et~al.}(2011)\citenamefont
  {Irastorza} \emph {et~al.}}]{Irastorza:2011gs}%
  \BibitemOpen
  \bibfield  {author} {\bibinfo {author} {\bibfnamefont {I.}~\bibnamefont
  {Irastorza}} \emph {et~al.} (\bibinfo {collaboration} {IAXO}),\ }\href
  {https://doi.org/10.1088/1475-7516/2011/06/013} {\bibfield  {journal}
  {\bibinfo  {journal} {JCAP}\ }\textbf {\bibinfo {volume} {2021}}\bibfield
  {number} {\bibinfo  {number} { (06)},\ \bibinfo {pages} {013}},\ }\Eprint
  {https://arxiv.org/abs/1103.5334} {arXiv:1103.5334 [hep-ex]} \BibitemShut
  {NoStop}%
\bibitem [{\citenamefont {{Esmat}}\ and\ \citenamefont {{Haloscope At Yale
  Sensitive To Axion Cdm (Haystac) Team}}(2022)}]{2022APS..APRS10006E}%
  \BibitemOpen
  \bibfield  {author} {\bibinfo {author} {\bibfnamefont {M.}~\bibnamefont
  {{Esmat}}}\ and\ \bibinfo {author} {\bibnamefont {{Haloscope At Yale
  Sensitive To Axion Cdm (Haystac) Team}}},\ }in\ \href@noop {} {\emph
  {\bibinfo {booktitle} {APS April Meeting Abstracts}}},\ \bibinfo {series}
  {APS Meeting Abstracts}, Vol.\ \bibinfo {volume} {2022}\ (\bibinfo {year}
  {2022})\ p.\ \bibinfo {pages} {S10.006}\BibitemShut {NoStop}%
\bibitem [{\citenamefont {Asztalos}\ \emph {et~al.}(2001)\citenamefont
  {Asztalos}, \citenamefont {Daw}, \citenamefont {Peng}, \citenamefont
  {Rosenberg}, \citenamefont {Hagmann}, \citenamefont {Kinion} \emph
  {et~al.}}]{PhysRevD.64.092003}%
  \BibitemOpen
  \bibfield  {author} {\bibinfo {author} {\bibfnamefont {S.}~\bibnamefont
  {Asztalos}}, \bibinfo {author} {\bibfnamefont {E.}~\bibnamefont {Daw}},
  \bibinfo {author} {\bibfnamefont {H.}~\bibnamefont {Peng}}, \bibinfo {author}
  {\bibfnamefont {L.~J.}\ \bibnamefont {Rosenberg}}, \bibinfo {author}
  {\bibfnamefont {C.}~\bibnamefont {Hagmann}}, \bibinfo {author} {\bibfnamefont
  {D.}~\bibnamefont {Kinion}}, \emph {et~al.},\ }\href
  {https://doi.org/10.1103/PhysRevD.64.092003} {\bibfield  {journal} {\bibinfo
  {journal} {Phys. Rev. D}\ }\textbf {\bibinfo {volume} {64}},\ \bibinfo
  {pages} {092003} (\bibinfo {year} {2001})}\BibitemShut {NoStop}%
\bibitem [{\citenamefont {Asztalos}\ \emph {et~al.}(2004)\citenamefont
  {Asztalos}, \citenamefont {Bradley}, \citenamefont {Duffy}, \citenamefont
  {Hagmann}, \citenamefont {Kinion}, \citenamefont {Moltz} \emph
  {et~al.}}]{PhysRevD.69.011101}%
  \BibitemOpen
  \bibfield  {author} {\bibinfo {author} {\bibfnamefont {S.}~\bibnamefont
  {Asztalos}}, \bibinfo {author} {\bibfnamefont {R.}~\bibnamefont {Bradley}},
  \bibinfo {author} {\bibfnamefont {L.}~\bibnamefont {Duffy}}, \bibinfo
  {author} {\bibfnamefont {C.}~\bibnamefont {Hagmann}}, \bibinfo {author}
  {\bibfnamefont {D.}~\bibnamefont {Kinion}}, \bibinfo {author} {\bibfnamefont
  {D.}~\bibnamefont {Moltz}}, \emph {et~al.},\ }\href
  {https://doi.org/10.1103/PhysRevD.69.011101} {\bibfield  {journal} {\bibinfo
  {journal} {Phys. Rev. D}\ }\textbf {\bibinfo {volume} {69}},\ \bibinfo
  {pages} {011101} (\bibinfo {year} {2004})}\BibitemShut {NoStop}%
\bibitem [{\citenamefont {Asztalos}\ \emph {et~al.}(2002)\citenamefont
  {Asztalos} \emph {et~al.}}]{Asztalos_2002}%
  \BibitemOpen
  \bibfield  {author} {\bibinfo {author} {\bibfnamefont {S.~J.}\ \bibnamefont
  {Asztalos}} \emph {et~al.},\ }\href {https://doi.org/10.1086/341130}
  {\bibfield  {journal} {\bibinfo  {journal} {The Astrophysical Journal}\
  }\textbf {\bibinfo {volume} {571}},\ \bibinfo {pages} {L27} (\bibinfo {year}
  {2002})}\BibitemShut {NoStop}%
\bibitem [{\citenamefont {Asztalos}\ \emph {et~al.}(2010)\citenamefont
  {Asztalos} \emph {et~al.}}]{PhysRevLett.104.041301}%
  \BibitemOpen
  \bibfield  {author} {\bibinfo {author} {\bibfnamefont {S.~J.}\ \bibnamefont
  {Asztalos}} \emph {et~al.},\ }\href
  {https://doi.org/10.1103/PhysRevLett.104.041301} {\bibfield  {journal}
  {\bibinfo  {journal} {Phys. Rev. Lett.}\ }\textbf {\bibinfo {volume} {104}},\
  \bibinfo {pages} {041301} (\bibinfo {year} {2010})}\BibitemShut {NoStop}%
\bibitem [{\citenamefont {Du}\ \emph {et~al.}(2018)\citenamefont {Du},
  \citenamefont {Force}, \citenamefont {Khatiwada}, \citenamefont {Lentz},
  \citenamefont {Ottens}, \citenamefont {Rosenberg} \emph
  {et~al.}}]{PhysRevLett.120.151301}%
  \BibitemOpen
  \bibfield  {author} {\bibinfo {author} {\bibfnamefont {N.}~\bibnamefont
  {Du}}, \bibinfo {author} {\bibfnamefont {N.}~\bibnamefont {Force}}, \bibinfo
  {author} {\bibfnamefont {R.}~\bibnamefont {Khatiwada}}, \bibinfo {author}
  {\bibfnamefont {E.}~\bibnamefont {Lentz}}, \bibinfo {author} {\bibfnamefont
  {R.}~\bibnamefont {Ottens}}, \bibinfo {author} {\bibfnamefont {L.~J.}\
  \bibnamefont {Rosenberg}}, \emph {et~al.} (\bibinfo {collaboration} {ADMX
  Collaboration}),\ }\href {https://doi.org/10.1103/PhysRevLett.120.151301}
  {\bibfield  {journal} {\bibinfo  {journal} {Phys. Rev. Lett.}\ }\textbf
  {\bibinfo {volume} {120}},\ \bibinfo {pages} {151301} (\bibinfo {year}
  {2018})}\BibitemShut {NoStop}%
\bibitem [{\citenamefont {Braine}\ \emph {et~al.}(2020)\citenamefont {Braine},
  \citenamefont {Cervantes}, \citenamefont {Crisosto}, \citenamefont {Du},
  \citenamefont {Kimes}, \citenamefont {Rosenberg} \emph
  {et~al.}}]{PhysRevLett.124.101303}%
  \BibitemOpen
  \bibfield  {author} {\bibinfo {author} {\bibfnamefont {T.}~\bibnamefont
  {Braine}}, \bibinfo {author} {\bibfnamefont {R.}~\bibnamefont {Cervantes}},
  \bibinfo {author} {\bibfnamefont {N.}~\bibnamefont {Crisosto}}, \bibinfo
  {author} {\bibfnamefont {N.}~\bibnamefont {Du}}, \bibinfo {author}
  {\bibfnamefont {S.}~\bibnamefont {Kimes}}, \bibinfo {author} {\bibfnamefont
  {L.~J.}\ \bibnamefont {Rosenberg}}, \emph {et~al.} (\bibinfo {collaboration}
  {ADMX Collaboration}),\ }\href
  {https://doi.org/10.1103/PhysRevLett.124.101303} {\bibfield  {journal}
  {\bibinfo  {journal} {Phys. Rev. Lett.}\ }\textbf {\bibinfo {volume} {124}},\
  \bibinfo {pages} {101303} (\bibinfo {year} {2020})}\BibitemShut {NoStop}%
\bibitem [{\citenamefont {Bartram}\ \emph {et~al.}(2021)\citenamefont {Bartram}
  \emph {et~al.}}]{PhysRevLett.127.261803}%
  \BibitemOpen
  \bibfield  {author} {\bibinfo {author} {\bibfnamefont {C.}~\bibnamefont
  {Bartram}} \emph {et~al.} (\bibinfo {collaboration} {ADMX Collaboration}),\
  }\href {https://doi.org/10.1103/PhysRevLett.127.261803} {\bibfield  {journal}
  {\bibinfo  {journal} {Phys. Rev. Lett.}\ }\textbf {\bibinfo {volume} {127}},\
  \bibinfo {pages} {261803} (\bibinfo {year} {2021})}\BibitemShut {NoStop}%
\bibitem [{Note1()}]{Note1}%
  \BibitemOpen
  \bibinfo {note} {For an alternative detection scheme which exploits the
  upcoming Square Kilometer Array to search for photon signals originating from
  the conversion of relativistic axions in astrophysical magnetic fields, see
  Ref.~\cite {kar2022searching}}\BibitemShut {NoStop}%
\bibitem [{\citenamefont {Elor}\ \emph {et~al.}(2015)\citenamefont {Elor},
  \citenamefont {Rodd},\ and\ \citenamefont {Slatyer}}]{Elor:2015tva}%
  \BibitemOpen
  \bibfield  {author} {\bibinfo {author} {\bibfnamefont {G.}~\bibnamefont
  {Elor}}, \bibinfo {author} {\bibfnamefont {N.~L.}\ \bibnamefont {Rodd}},\
  and\ \bibinfo {author} {\bibfnamefont {T.~R.}\ \bibnamefont {Slatyer}},\
  }\href {https://doi.org/10.1103/PhysRevD.91.103531} {\bibfield  {journal}
  {\bibinfo  {journal} {Phys. Rev. D}\ }\textbf {\bibinfo {volume} {91}},\
  \bibinfo {pages} {103531} (\bibinfo {year} {2015})},\ \Eprint
  {https://arxiv.org/abs/1503.01773} {arXiv:1503.01773 [hep-ph]} \BibitemShut
  {NoStop}%
\bibitem [{\citenamefont {Elor}\ \emph {et~al.}(2016)\citenamefont {Elor},
  \citenamefont {Rodd}, \citenamefont {Slatyer},\ and\ \citenamefont
  {Xue}}]{Elor:2015bho}%
  \BibitemOpen
  \bibfield  {author} {\bibinfo {author} {\bibfnamefont {G.}~\bibnamefont
  {Elor}}, \bibinfo {author} {\bibfnamefont {N.~L.}\ \bibnamefont {Rodd}},
  \bibinfo {author} {\bibfnamefont {T.~R.}\ \bibnamefont {Slatyer}},\ and\
  \bibinfo {author} {\bibfnamefont {W.}~\bibnamefont {Xue}},\ }\href
  {https://doi.org/10.1088/1475-7516/2016/06/024} {\bibfield  {journal}
  {\bibinfo  {journal} {JCAP}\ }\textbf {\bibinfo {volume} {06}},\ \bibinfo
  {pages} {024}},\ \Eprint {https://arxiv.org/abs/1511.08787} {arXiv:1511.08787
  [hep-ph]} \BibitemShut {NoStop}%
\bibitem [{\citenamefont {Chen}\ \emph {et~al.}(2021)\citenamefont {Chen} \emph
  {et~al.}}]{PhysRevD.103.123528}%
  \BibitemOpen
  \bibfield  {author} {\bibinfo {author} {\bibfnamefont {A.}~\bibnamefont
  {Chen}} \emph {et~al.},\ }\href {https://doi.org/10.1103/PhysRevD.103.123528}
  {\bibfield  {journal} {\bibinfo  {journal} {Phys. Rev. D}\ }\textbf {\bibinfo
  {volume} {103}},\ \bibinfo {pages} {123528} (\bibinfo {year}
  {2021})}\BibitemShut {NoStop}%
\bibitem [{Note2()}]{Note2}%
  \BibitemOpen
  \bibinfo {note} {To derive this value from Ref.~\cite {PhysRevD.103.123528},
  we set the ratio of the amount of dark matter converted into dark radiation
  divided by the amount of dark matter today, which we approximate by
  $(1-e^{-t_{\scriptscriptstyle {\protect \rm U}}/\tau
  })/e^{-t_{\scriptscriptstyle {\protect \rm U}}/\tau }$, equal to their 95\%
  C.L. limit on $\zeta $, marginalized over all nuisance parameters
  (0.72)}\BibitemShut {NoStop}%
\bibitem [{Note3()}]{Note3}%
  \BibitemOpen
  \bibinfo {note} {We emphasize that the daily modulation signal we describe
  arises solely from the interaction of the \protect \ensuremath {{\protect \rm
  C}a{\protect \rm B}}\protect \xspace wave with the cavity. This can be
  contrasted with other possible modulating signals, such as that proposed for
  Axion Quark Nugget dark matter, which generates a relativistic axion signal
  where the amplitude of the signal oscillates throughout the day, see
  Refs.~\cite {Fischer:2018niu,Liang:2019lya,Budker:2019zka}.}\BibitemShut
  {Stop}%
\bibitem [{\citenamefont {Siddiqi}\ \emph {et~al.}(2004)\citenamefont
  {Siddiqi}, \citenamefont {Vijay}, \citenamefont {Pierre}, \citenamefont
  {Wilson}, \citenamefont {Metcalfe}, \citenamefont {Rigetti}, \citenamefont
  {Frunzio},\ and\ \citenamefont {Devoret}}]{Siddiqi_2004}%
  \BibitemOpen
  \bibfield  {author} {\bibinfo {author} {\bibfnamefont {I.}~\bibnamefont
  {Siddiqi}}, \bibinfo {author} {\bibfnamefont {R.}~\bibnamefont {Vijay}},
  \bibinfo {author} {\bibfnamefont {F.}~\bibnamefont {Pierre}}, \bibinfo
  {author} {\bibfnamefont {C.~M.}\ \bibnamefont {Wilson}}, \bibinfo {author}
  {\bibfnamefont {M.}~\bibnamefont {Metcalfe}}, \bibinfo {author}
  {\bibfnamefont {C.}~\bibnamefont {Rigetti}}, \bibinfo {author} {\bibfnamefont
  {L.}~\bibnamefont {Frunzio}},\ and\ \bibinfo {author} {\bibfnamefont {M.~H.}\
  \bibnamefont {Devoret}},\ }\href
  {https://doi.org/10.1103/physrevlett.93.207002} {\bibfield  {journal}
  {\bibinfo  {journal} {Physical Review Letters}\ }\textbf {\bibinfo {volume}
  {93}},\ \bibinfo {pages} {207002} (\bibinfo {year} {2004})}\BibitemShut
  {NoStop}%
\bibitem [{\citenamefont {Factory}()}]{LNFLNC026A}%
  \BibitemOpen
  \bibfield  {author} {\bibinfo {author} {\bibfnamefont {L.~N.}\ \bibnamefont
  {Factory}},\ }\href@noop {} {\bibinfo {title} {0.6-2 ghz cryogenic low noise
  amplifier}},\ \bibinfo {howpublished}
  {\url{https://www.lownoisefactory.com/files/1215/2585/7504/LNF-LNC0.6_2A.pdf}}\BibitemShut
  {NoStop}%
\bibitem [{\citenamefont {Khatiwada}\ \emph {et~al.}(2021)\citenamefont
  {Khatiwada} \emph {et~al.}}]{doi:10.1063/5.0037857}%
  \BibitemOpen
  \bibfield  {author} {\bibinfo {author} {\bibfnamefont {R.}~\bibnamefont
  {Khatiwada}} \emph {et~al.},\ }\href {https://doi.org/10.1063/5.0037857}
  {\bibfield  {journal} {\bibinfo  {journal} {Review of Scientific
  Instruments}\ }\textbf {\bibinfo {volume} {92}},\ \bibinfo {pages} {124502}
  (\bibinfo {year} {2021})},\ \Eprint
  {https://arxiv.org/abs/https://doi.org/10.1063/5.0037857}
  {https://doi.org/10.1063/5.0037857} \BibitemShut {NoStop}%
\bibitem [{Note4()}]{Note4}%
  \BibitemOpen
  \bibinfo {note} {Representative uncertainties are related to electric field
  distribution, noise temperature measurements, and cavity parameters, see Ref.
  \cite {PhysRevLett.127.261803}.}\BibitemShut {Stop}%
\bibitem [{\citenamefont {Anastassopoulos}\ \emph {et~al.}(2017)\citenamefont
  {Anastassopoulos} \emph {et~al.}}]{Anastassopoulos2017}%
  \BibitemOpen
  \bibfield  {author} {\bibinfo {author} {\bibfnamefont {V.}~\bibnamefont
  {Anastassopoulos}} \emph {et~al.},\ }\href
  {https://doi.org/10.1038/nphys4109} {\bibfield  {journal} {\bibinfo
  {journal} {Nature Physics}\ }\textbf {\bibinfo {volume} {13}},\ \bibinfo
  {pages} {584} (\bibinfo {year} {2017})}\BibitemShut {NoStop}%
\bibitem [{\citenamefont {Dixit}\ \emph {et~al.}(2021)\citenamefont {Dixit},
  \citenamefont {Chakram}, \citenamefont {He}, \citenamefont {Agrawal},
  \citenamefont {Naik}, \citenamefont {Schuster},\ and\ \citenamefont
  {Chou}}]{PhysRevLett.126.141302}%
  \BibitemOpen
  \bibfield  {author} {\bibinfo {author} {\bibfnamefont {A.~V.}\ \bibnamefont
  {Dixit}}, \bibinfo {author} {\bibfnamefont {S.}~\bibnamefont {Chakram}},
  \bibinfo {author} {\bibfnamefont {K.}~\bibnamefont {He}}, \bibinfo {author}
  {\bibfnamefont {A.}~\bibnamefont {Agrawal}}, \bibinfo {author} {\bibfnamefont
  {R.~K.}\ \bibnamefont {Naik}}, \bibinfo {author} {\bibfnamefont {D.~I.}\
  \bibnamefont {Schuster}},\ and\ \bibinfo {author} {\bibfnamefont
  {A.}~\bibnamefont {Chou}},\ }\href
  {https://doi.org/10.1103/PhysRevLett.126.141302} {\bibfield  {journal}
  {\bibinfo  {journal} {Phys. Rev. Lett.}\ }\textbf {\bibinfo {volume} {126}},\
  \bibinfo {pages} {141302} (\bibinfo {year} {2021})}\BibitemShut {NoStop}%
\bibitem [{\citenamefont {Brune}\ \emph {et~al.}(1990)\citenamefont {Brune},
  \citenamefont {Haroche}, \citenamefont {Lefevre}, \citenamefont {Raimond},\
  and\ \citenamefont {Zagury}}]{PhysRevLett.65.976}%
  \BibitemOpen
  \bibfield  {author} {\bibinfo {author} {\bibfnamefont {M.}~\bibnamefont
  {Brune}}, \bibinfo {author} {\bibfnamefont {S.}~\bibnamefont {Haroche}},
  \bibinfo {author} {\bibfnamefont {V.}~\bibnamefont {Lefevre}}, \bibinfo
  {author} {\bibfnamefont {J.~M.}\ \bibnamefont {Raimond}},\ and\ \bibinfo
  {author} {\bibfnamefont {N.}~\bibnamefont {Zagury}},\ }\href
  {https://doi.org/10.1103/PhysRevLett.65.976} {\bibfield  {journal} {\bibinfo
  {journal} {Phys. Rev. Lett.}\ }\textbf {\bibinfo {volume} {65}},\ \bibinfo
  {pages} {976} (\bibinfo {year} {1990})}\BibitemShut {NoStop}%
\bibitem [{\citenamefont {Berlin}\ \emph {et~al.}(2021)\citenamefont {Berlin},
  \citenamefont {Blas}, \citenamefont {D'Agnolo}, \citenamefont {Ellis},
  \citenamefont {Harnik}, \citenamefont {Kahn},\ and\ \citenamefont
  {Schütte-Engel}}]{berlin2021detecting}%
  \BibitemOpen
  \bibfield  {author} {\bibinfo {author} {\bibfnamefont {A.}~\bibnamefont
  {Berlin}}, \bibinfo {author} {\bibfnamefont {D.}~\bibnamefont {Blas}},
  \bibinfo {author} {\bibfnamefont {R.~T.}\ \bibnamefont {D'Agnolo}}, \bibinfo
  {author} {\bibfnamefont {S.~A.~R.}\ \bibnamefont {Ellis}}, \bibinfo {author}
  {\bibfnamefont {R.}~\bibnamefont {Harnik}}, \bibinfo {author} {\bibfnamefont
  {Y.}~\bibnamefont {Kahn}},\ and\ \bibinfo {author} {\bibfnamefont
  {J.}~\bibnamefont {Schütte-Engel}},\ }\href@noop {} {\bibinfo {title}
  {Detecting high-frequency gravitational waves with microwave cavities}}
  (\bibinfo {year} {2021}),\ \Eprint {https://arxiv.org/abs/2112.11465}
  {arXiv:2112.11465 [hep-ph]} \BibitemShut {NoStop}%
\bibitem [{\citenamefont {Domcke}\ \emph {et~al.}(2022)\citenamefont {Domcke},
  \citenamefont {Garcia-Cely},\ and\ \citenamefont {Rodd}}]{Domcke:2022rgu}%
  \BibitemOpen
  \bibfield  {author} {\bibinfo {author} {\bibfnamefont {V.}~\bibnamefont
  {Domcke}}, \bibinfo {author} {\bibfnamefont {C.}~\bibnamefont
  {Garcia-Cely}},\ and\ \bibinfo {author} {\bibfnamefont {N.~L.}\ \bibnamefont
  {Rodd}},\ }\href {https://doi.org/10.1103/PhysRevLett.129.041101} {\bibfield
  {journal} {\bibinfo  {journal} {Phys. Rev. Lett.}\ }\textbf {\bibinfo
  {volume} {129}},\ \bibinfo {pages} {041101} (\bibinfo {year} {2022})},\
  \Eprint {https://arxiv.org/abs/2202.00695} {arXiv:2202.00695 [hep-ph]}
  \BibitemShut {NoStop}%
\bibitem [{\citenamefont {Kar}\ \emph {et~al.}(2022)\citenamefont {Kar},
  \citenamefont {Kumar}, \citenamefont {Roy},\ and\ \citenamefont
  {Zupan}}]{kar2022searching}%
  \BibitemOpen
  \bibfield  {author} {\bibinfo {author} {\bibfnamefont {A.}~\bibnamefont
  {Kar}}, \bibinfo {author} {\bibfnamefont {T.}~\bibnamefont {Kumar}}, \bibinfo
  {author} {\bibfnamefont {S.}~\bibnamefont {Roy}},\ and\ \bibinfo {author}
  {\bibfnamefont {J.}~\bibnamefont {Zupan}},\ }\href@noop {} {\bibinfo {title}
  {Searching for relativistic axions in the sky}} (\bibinfo {year} {2022}),\
  \Eprint {https://arxiv.org/abs/2212.04647} {arXiv:2212.04647 [hep-ph]}
  \BibitemShut {NoStop}%
\bibitem [{\citenamefont {Fischer}\ \emph {et~al.}(2018)\citenamefont
  {Fischer}, \citenamefont {Liang}, \citenamefont {Semertzidis}, \citenamefont
  {Zhitnitsky},\ and\ \citenamefont {Zioutas}}]{Fischer:2018niu}%
  \BibitemOpen
  \bibfield  {author} {\bibinfo {author} {\bibfnamefont {H.}~\bibnamefont
  {Fischer}}, \bibinfo {author} {\bibfnamefont {X.}~\bibnamefont {Liang}},
  \bibinfo {author} {\bibfnamefont {Y.}~\bibnamefont {Semertzidis}}, \bibinfo
  {author} {\bibfnamefont {A.}~\bibnamefont {Zhitnitsky}},\ and\ \bibinfo
  {author} {\bibfnamefont {K.}~\bibnamefont {Zioutas}},\ }\href
  {https://doi.org/10.1103/PhysRevD.98.043013} {\bibfield  {journal} {\bibinfo
  {journal} {Phys. Rev. D}\ }\textbf {\bibinfo {volume} {98}},\ \bibinfo
  {pages} {043013} (\bibinfo {year} {2018})},\ \Eprint
  {https://arxiv.org/abs/1805.05184} {arXiv:1805.05184 [hep-ph]} \BibitemShut
  {NoStop}%
\bibitem [{\citenamefont {Liang}\ \emph {et~al.}(2020)\citenamefont {Liang},
  \citenamefont {Mead}, \citenamefont {Siddiqui}, \citenamefont
  {Van~Waerbeke},\ and\ \citenamefont {Zhitnitsky}}]{Liang:2019lya}%
  \BibitemOpen
  \bibfield  {author} {\bibinfo {author} {\bibfnamefont {X.}~\bibnamefont
  {Liang}}, \bibinfo {author} {\bibfnamefont {A.}~\bibnamefont {Mead}},
  \bibinfo {author} {\bibfnamefont {M.~S.~R.}\ \bibnamefont {Siddiqui}},
  \bibinfo {author} {\bibfnamefont {L.}~\bibnamefont {Van~Waerbeke}},\ and\
  \bibinfo {author} {\bibfnamefont {A.}~\bibnamefont {Zhitnitsky}},\ }\href
  {https://doi.org/10.1103/PhysRevD.101.043512} {\bibfield  {journal} {\bibinfo
   {journal} {Phys. Rev. D}\ }\textbf {\bibinfo {volume} {101}},\ \bibinfo
  {pages} {043512} (\bibinfo {year} {2020})},\ \Eprint
  {https://arxiv.org/abs/1908.04675} {arXiv:1908.04675 [astro-ph.CO]}
  \BibitemShut {NoStop}%
\bibitem [{\citenamefont {Budker}\ \emph {et~al.}(2020)\citenamefont {Budker},
  \citenamefont {Flambaum}, \citenamefont {Liang},\ and\ \citenamefont
  {Zhitnitsky}}]{Budker:2019zka}%
  \BibitemOpen
  \bibfield  {author} {\bibinfo {author} {\bibfnamefont {D.}~\bibnamefont
  {Budker}}, \bibinfo {author} {\bibfnamefont {V.~V.}\ \bibnamefont
  {Flambaum}}, \bibinfo {author} {\bibfnamefont {X.}~\bibnamefont {Liang}},\
  and\ \bibinfo {author} {\bibfnamefont {A.}~\bibnamefont {Zhitnitsky}},\
  }\href {https://doi.org/10.1103/PhysRevD.101.043012} {\bibfield  {journal}
  {\bibinfo  {journal} {Phys. Rev. D}\ }\textbf {\bibinfo {volume} {101}},\
  \bibinfo {pages} {043012} (\bibinfo {year} {2020})},\ \Eprint
  {https://arxiv.org/abs/1909.09475} {arXiv:1909.09475 [hep-ph]} \BibitemShut
  {NoStop}%
\bibitem [{\citenamefont {Navarro}\ \emph {et~al.}(1996)\citenamefont
  {Navarro}, \citenamefont {Frenk},\ and\ \citenamefont
  {White}}]{Navarro:1995iw}%
  \BibitemOpen
  \bibfield  {author} {\bibinfo {author} {\bibfnamefont {J.~F.}\ \bibnamefont
  {Navarro}}, \bibinfo {author} {\bibfnamefont {C.~S.}\ \bibnamefont {Frenk}},\
  and\ \bibinfo {author} {\bibfnamefont {S.~D.~M.}\ \bibnamefont {White}},\
  }\href {https://doi.org/10.1086/177173} {\bibfield  {journal} {\bibinfo
  {journal} {Astrophys. J.}\ }\textbf {\bibinfo {volume} {462}},\ \bibinfo
  {pages} {563} (\bibinfo {year} {1996})},\ \Eprint
  {https://arxiv.org/abs/astro-ph/9508025} {arXiv:astro-ph/9508025}
  \BibitemShut {NoStop}%
\bibitem [{\citenamefont {Navarro}\ \emph {et~al.}(1997)\citenamefont
  {Navarro}, \citenamefont {Frenk},\ and\ \citenamefont
  {White}}]{Navarro:1996gj}%
  \BibitemOpen
  \bibfield  {author} {\bibinfo {author} {\bibfnamefont {J.~F.}\ \bibnamefont
  {Navarro}}, \bibinfo {author} {\bibfnamefont {C.~S.}\ \bibnamefont {Frenk}},\
  and\ \bibinfo {author} {\bibfnamefont {S.~D.~M.}\ \bibnamefont {White}},\
  }\href {https://doi.org/10.1086/304888} {\bibfield  {journal} {\bibinfo
  {journal} {Astrophys. J.}\ }\textbf {\bibinfo {volume} {490}},\ \bibinfo
  {pages} {493} (\bibinfo {year} {1997})},\ \Eprint
  {https://arxiv.org/abs/astro-ph/9611107} {arXiv:astro-ph/9611107}
  \BibitemShut {NoStop}%
\end{thebibliography}%

\clearpage
\onecolumngrid
\begin{center}
   \textbf{\large SUPPLEMENTARY MATERIAL \\[.2cm] ``Search for a dark-matter induced Cosmic Axion Background with ADMX''}\\[.2cm]
\end{center}

\setcounter{equation}{0}
\setcounter{figure}{0}
\setcounter{table}{0}
\setcounter{section}{0}
\setcounter{page}{1}
\makeatletter
\renewcommand{\theequation}{S\arabic{equation}}
\renewcommand{\thefigure}{S\arabic{figure}}
\renewcommand{\thetable}{S\arabic{table}}

\onecolumngrid

In this supplementary material we provide additional details regarding three points from the main text.
Firstly, we describe the dark matter decay model we study to generate the \CaB, and provide the full details of Eq.~\eqref{eq:dPdw}.
Secondly, we describe the origin of the suppression of C$a$B power that results from the form factor in Eq.~\eqref{eq:K}.
Finally, the scenarios we consider depend on three parameters: the dark-matter mass, the lifetime for the decay, and the axion-photon coupling.
In the main text we fixed the lifetime and studied constraints as a function of the remaining two parameters, and here we study the impact of varying $\tau$.

\section*{The Dark Matter Model}

We begin with the dark matter model.
In a scenario where light dark matter decays to axions, we emphasize that a direct decay $\chi \to aa$ can rapidly deplete all of the dark matter.
To see this, note that the occupation number of final state axions as a function of their energy $\omega$ is given by
\begin{equation}
f_a(\omega) = \frac{2\pi^2}{\omega^3} \frac{d\rho_a}{d\omega}.
\end{equation}
This is a general result.
If we apply it to the case of a two body decay, then the spectrum peaks at $\omega = m_{\rm DM}/2$, and we find approximately 
\begin{equation}
f_a(m_{\rm DM}/2) \sim 10^9 \left( \frac{\rho_a}{\rho_\gamma} \right) \left( \frac{m_{\rm DM}}{10~\mu{\rm eV}} \right)^{-4}\!.
\label{eq:fa}
\end{equation}
Accordingly, in the range of mass to which ADMX is sensitive, if we want an appreciable \CaB density (for instance, $\rho_a \gtrsim \rho_{\gamma}$), the occupation number will be enormous.
As axions are bosons, this large occupation number will significantly bose enhance the decay of dark matter itself, leading to a runaway process that will deplete dark matter.

This observation motivated our consideration of the next-to-minimal scenario involving a cascade decay through a mediator, $\varphi$.
In particular, in the main text, we studied $\chi \to \varphi \varphi \to a a a a$.
Here we present a summary of a simple benchmark model which exhibits cascade decays and whose spectrum is used as the template in our experiment.
The model assumes that dark matter and the mediator are real scalar fields and with interactions set by
\begin{equation} 
{\cal L} \supset  \frac{1}{2} g \chi \varphi^2 + \frac{1}{2\Lambda} \varphi (\partial_\mu a)^2,
\end{equation}
where $g$ is a dimensioned coupling.
Note that radiative corrections generate a $\chi (\partial a)^2$ coupling and hence a direct $\chi \rightarrow a a $ decay. Relative to the $\chi \rightarrow  \varphi \varphi $ decay rate, shown below in Eq.~\eqref{eq:chiphiphi}, the $\chi \rightarrow a a$ rate is suppressed by $(m_{\rm DM}/\Lambda)^4$.
If the product of this factor and the Bose-enhancement factor in Eq.~\eqref{eq:fa} is small, $\chi\rightarrow a a $ is slow on cosmic timescales and we assume this condition is satisfied.
We note that this condition is non-trivial to satisfy since we also assume that the decay of $\varphi$, controlled by $\Lambda$, is sufficiently fast that it occurs promptly over astrophysical scales.
Nonetheless, there is ample parameter space where both requirements can be met.

In this model, the lifetime of dark matter is controlled by the initial decay, and for $m_\varphi \ll m_{\rm DM}$ given by 
\begin{equation} 
\Gamma_{\chi \to \varphi \varphi} = \frac{g^2}{16 \pi m_{\rm DM}} \simeq \frac{1}{10\,\tU} \left( \frac{g}{10^{-19}~{\rm eV}} \right)^2   \left( \frac{m_{\rm DM}}{1~\mu{\rm eV}} \right)^{-1}\!.
\label{eq:chiphiphi}
\end{equation}
To produce a detectable flux of axions, we want as large a decay rate as possible, although we still require $\Gamma^{-1} = \tau \lesssim 1.8\,\tU$ as measured by DES. We see that in order for this model to yield a signal consistent with observations and be detectable by ADMX requires a tiny value of the coupling $g$.
An additional simplification that occurs in the limit of $m_\varphi \ll m_{\rm DM}$ is that the energy spectrum of the axions emitted for each decay takes on a particularly simple form (see, e.g., Refs.~\cite{Elor:2015tva,Elor:2015bho})
\begin{equation} 
\frac{dN}{d \omega} = \frac{8}{m_{\rm DM}} \Theta(m_{\rm DM}/2-\omega),   
\end{equation}
where $\Theta$ denotes the Heaviside step function.
Importantly, as this depends only on $m_{\rm DM}$ in this limit, other models that satisfy these kinematic conditions would generate an equivalent spectrum.

In the cascade scenarios we consider, there are three relevant parameters: $(m_{\rm DM},\,\tau,\,\ga)$.
In Fig.~\ref{fig:limits} we chose to fix $\tau$, and then present results in the $(m_{\rm DM},\,\ga)$ plane.
For the scenario above, that corresponds to fixing $g$; for example, at $m_{\rm DM} = 10~\mu{\rm eV}$ we are fixing $g\simeq 6.5 \times 10^{-19}~{\rm eV}$.
If we had instead fixed $\ga$, we could interpret our results purely as a constraint on $g$ for this explicit scenario.
As a single example, at $10~\mu{\rm eV}$, our constraint in the main body on $\ga$ is roughly fifty times what CAST has obtained.
If we instead fix $\ga = 10^3 \gaCAST$, then our results require $\tau \gtrsim 1600\,\tU$, and therefore $g \lesssim 2 \times 10^{-20}~{\rm eV}$.

\section*{Complete Expressions for the Dark Matter Power}

From the lifetime and spectrum, we can compute the energy density of axions produced locally, and hence the differential power generated in a resonant cavity.
This was the result given in Eq.~\eqref{eq:dPdw}.
Most of the contributions to that expression were given in the main text, however, we did not state $\Omega_a^\MW(\omega)$, $\Omega_a^\EG(\omega)$, or $D_{\nu}(z)$ explicitly, and do so now.
Firstly, assuming an NFW dark matter distribution in the MW~\cite{Navarro:1995iw,Navarro:1996gj}, we obtain the following local density of axions from dark matter decay
\begin{equation}
\Omega_a^\MW(\omega) = \frac{2 e^{-\tU/\tau}}{\pi \tau \rho_c} \frac{\omega^2}{m_{\rm DM}^2} \frac{4\pi \rho_0 r_s \ln \nu}{1-\nu^{-2}}\, \Theta(m_{\rm DM}/2-\omega).
\end{equation}
Here $\rho_0 \simeq 0.32~{\rm GeV/cm}^3$ and $\nu = r_s/r_{\odot}$ determine the NFW profile, with $r_s = 20~{\rm kpc}$ the scale radius and $r_{\odot}$ the distance of the Sun from the Galactic Center.
Note that due to the exponential depletion of dark matter locally for $\tau \ll \tU$, the axion density does not simply scale as $1/\tau$ (although it does when $\tau \gg \tU$).
The extragalactic decays also contribute a density of axions, given by
\begin{equation}
\Omega_a^\EG(\omega) = \frac{8\Omega_{\scriptscriptstyle \textrm{DM}} \omega^2}{\tau H_0 m_{\rm DM}^2} \int_0^1 \frac{da}{a} \frac{e^{-(t(a)-t_r)/\tau}}{\sqrt{\Omega_m/a^3+\Omega_{\Lambda}}} \Theta(a-2\omega/m_{\rm DM}).
\end{equation}
Here $H_0$ is the Hubble constant, while $\Omega_{\scriptscriptstyle \textrm{DM}}$, $\Omega_m$, and $\Omega_{\Lambda}$ are the cosmological dark matter, matter, and dark energy densities, respectively.
The integral is performed over the scale factor $a$, and $t(a)$ is the age of the universe at that scale factor, so that $t(1) = \tU$.
Finally, in a model where dark matter can decay, we need to normalize $\Omega_{\scriptscriptstyle \textrm{DM}}$ to a point in time when it has been measured, and we have chosen to do so at $t_r$, the time of recombination.
Again this expression has a complicated dependence on the dark matter lifetime, except in the limit where $\tau \gg \tU$.

The density of axions produced from decays within and outside our own Galaxy are comparable.
Nevertheless, the EG contribution is essentially isotropic in the sky and therefore does not contribute significantly to the PEM.
By isolating time-varying signals, the PEM singles out the local decays, which primarily originate from the Galactic Center.
Accordingly, we need to know the angular dependence of the incident axions, and this is determined by ${\cal D}_{\nu}(z)$, as described in the main text.
If we again assume an NFW profile, then we can explicitly evaluate
\begin{equation}\begin{aligned}
{\cal D}_{\nu}(z) &= \frac{\nu-1}{4\ln \nu (\nu^2+z^2-1)^{3/2}} \left[
\nu(\nu+1) \operatorname{atanh} \left( \frac{2z \nu \sqrt{\nu^2+z^2-1}}{\nu^2+z^2(\nu^2+1)-1} \right) \right.\\
&\hspace{4cm}-2\nu(\nu+1) \operatorname{atanh} \left( \frac{\nu^2 - z \nu + z^2 - 1}{(z-\nu)\sqrt{\nu^2+z^2-1}} \right) \\
&\hspace{4cm}-\left. 2 (\nu+z+1) \sqrt{\nu^2+z^2-1}
\right]\!.
\end{aligned}\end{equation}

\section*{Suppression from the Relativistic Form Factor}

Finally, we expand on the physical origin of the suppression that arises from the relativistic form factor $K(\omega,\alpha)$.
In the main text, and particularly in Fig.~\ref{fig:CaBff}, we emphasized that the origin of this suppression is that the C$a$B spatially oscillates over the cavity, and when integrated over the magnetic volume, the net effect is a suppressed total axion power compared to dark matter, which is spatially coherent over the instrument.
Of course, we note that although this effect suppresses the total power, it also gave rise to the daily modulation we exploited in the PEM analysis.

\begin{figure}[!t]
\centering
\includegraphics[width=0.45\linewidth]{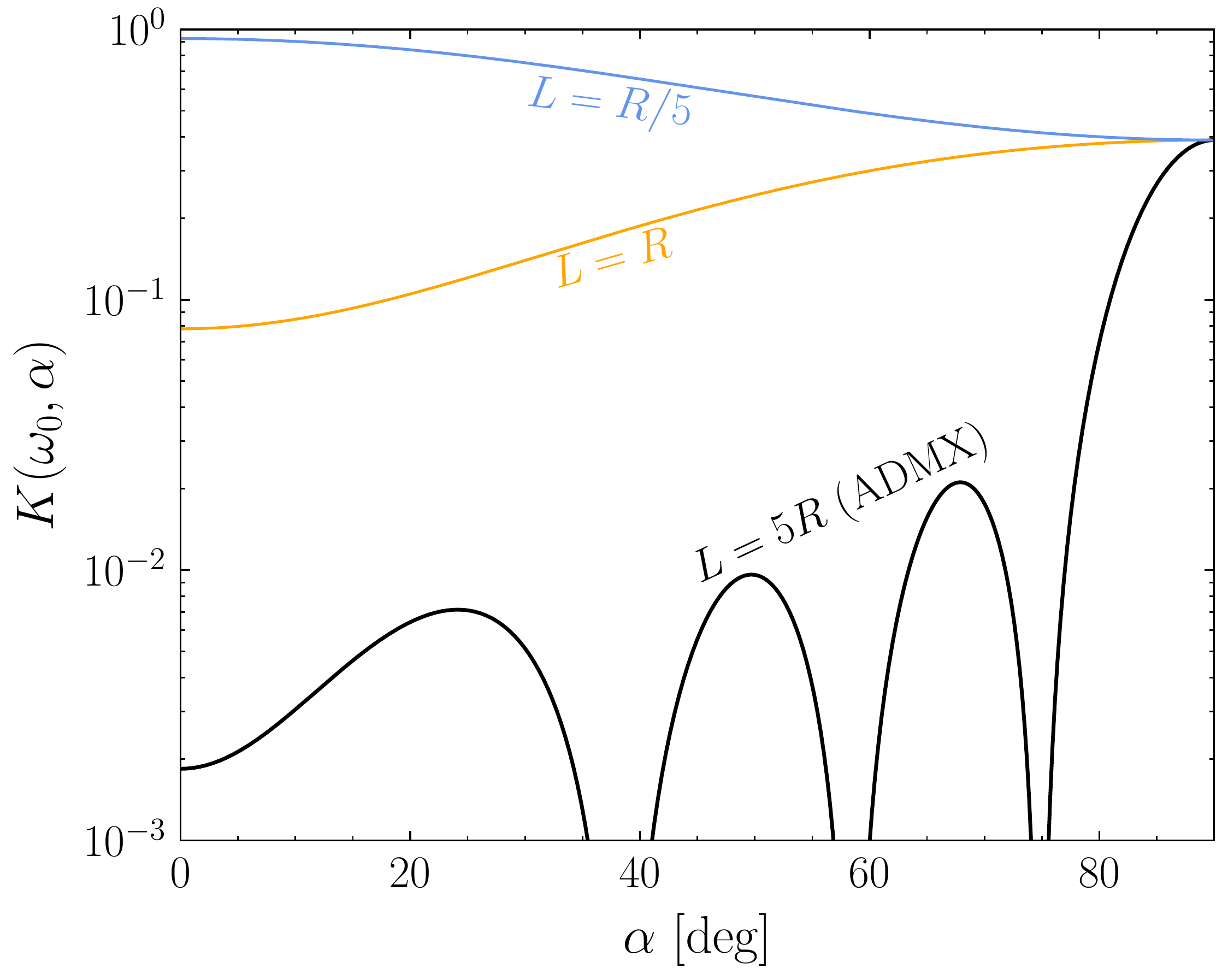}
\caption{The relativistic cavity form factor from Eq.~\eqref{eq:K}, evaluated at the nominal TM$_{010}$ resonant frequency $\omega_0 = j_{01}/R$, for different values of the ratio $L/R$.
The ADMX geometry, where $L=5R$, was shown as representing the C$a$B in Fig.~\ref{fig:CaBff}, and contrasted with the equivalent dark matter value of $K(\omega_0,\alpha)=1$.
As can be seen, when $L \gg R$ (as for ADMX), there is a significant suppression of the form factor.
For $L = R$, or $L \ll R$, the effect is greatly reduced, and the form factor becomes comparable to dark matter.
The origin of this behavior, and the consequences for C$a$B searches are discussed in the text.
}
\label{fig:relK}
\end{figure}

While the above discussion is correct, it may give the impression that such a suppression of power is an intrinsic property of relativistic axions detected in resonant cavities, and this is not correct.
This point is emphasized in Fig.~\ref{fig:relK}, where we show $K(\omega,\alpha)$ evaluated at the resonant frequency for three choices of the ratio $L/R$.
From there, we can see that in a cavity with $L=R$, or especially if $L \ll R$, the form factor is significantly larger.
This will reduce the size of any daily modulation signal, but once ADMX can make absolute power measurements, such a geometry would lead to a larger C$a$B signal, as was shown in Fig.~\ref{fig:limits}.

The physical origin of this is that the resonant frequency for the TM$_{010}$ mode of an idealized cylindrical cavity is given by $\omega_0 = j_{01}/R$.
Importantly, $R$ not $L$ determines the resonant frequency.
As the cavity transfer function only has appreciable support for $\omega \simeq \omega_0$ [see Eq.~\eqref{eq:dPdw}], this implies that C$a$B modes with wavelengths $\sim$$R$ are those which primarily source power.
If $L \ll R$, then for any incident direction, the primary modes have wavelengths that do not oscillate significantly over the cavity.
If $L \gg R$, however, then unless $\alpha \simeq 90^{\circ}$ (when the C$a$B wave is incident perpendicular to the magnetic field), the wave will oscillate for several cycles, leading to a suppression.

We can also see this analytically.
On resonance,
\begin{equation}
K(\omega_0,\alpha) = \left[ \frac{\sin(j_{01} \cos \alpha L/R)}{j_{01} \cos \alpha L/R} \frac{J_0(j_{01} \sin \alpha)}{1-\sin^2 \alpha} \right]^2\!.
\end{equation}
Firstly, when the incident wave and magnetic field are perpendicular, we have $K(\omega_0,\pi/2) = [j_{01} J_1(j_{01})]^2/4 \simeq 0.39$.
This result is independent of $L/R$, because the wave is propagating radially across the cavity, and the distance travelled and resonant mode are both controlled by $R$.
On the other hand, for $\alpha=0$, we have
\begin{equation}
K(\omega_0,0) = \left[ \frac{\sin(j_{01} L/R)}{j_{01} L/R}\right]^2\!.
\end{equation}
For $L/R \to 0$, this result tends to unity -- exactly reproducing the dark matter result.
For $L/R \to \infty$, however, the field oscillates many times across the cavity, and the form factor is driven to zero.
For the ADMX geometry, where $L=5R$, there is already significant suppression.
Of course, we emphasize that in the model we study, the axions are incident from across the sky, and therefore we are sensitive to a weighted integral of the form factor over all angles.
Nevertheless, with $L=R$, the integrated power from the C$a$B would be increased by roughly a factor of five.

\section*{Results when varying the lifetime}

\begin{figure}[!t]
\centering
\includegraphics[width=0.45\linewidth]{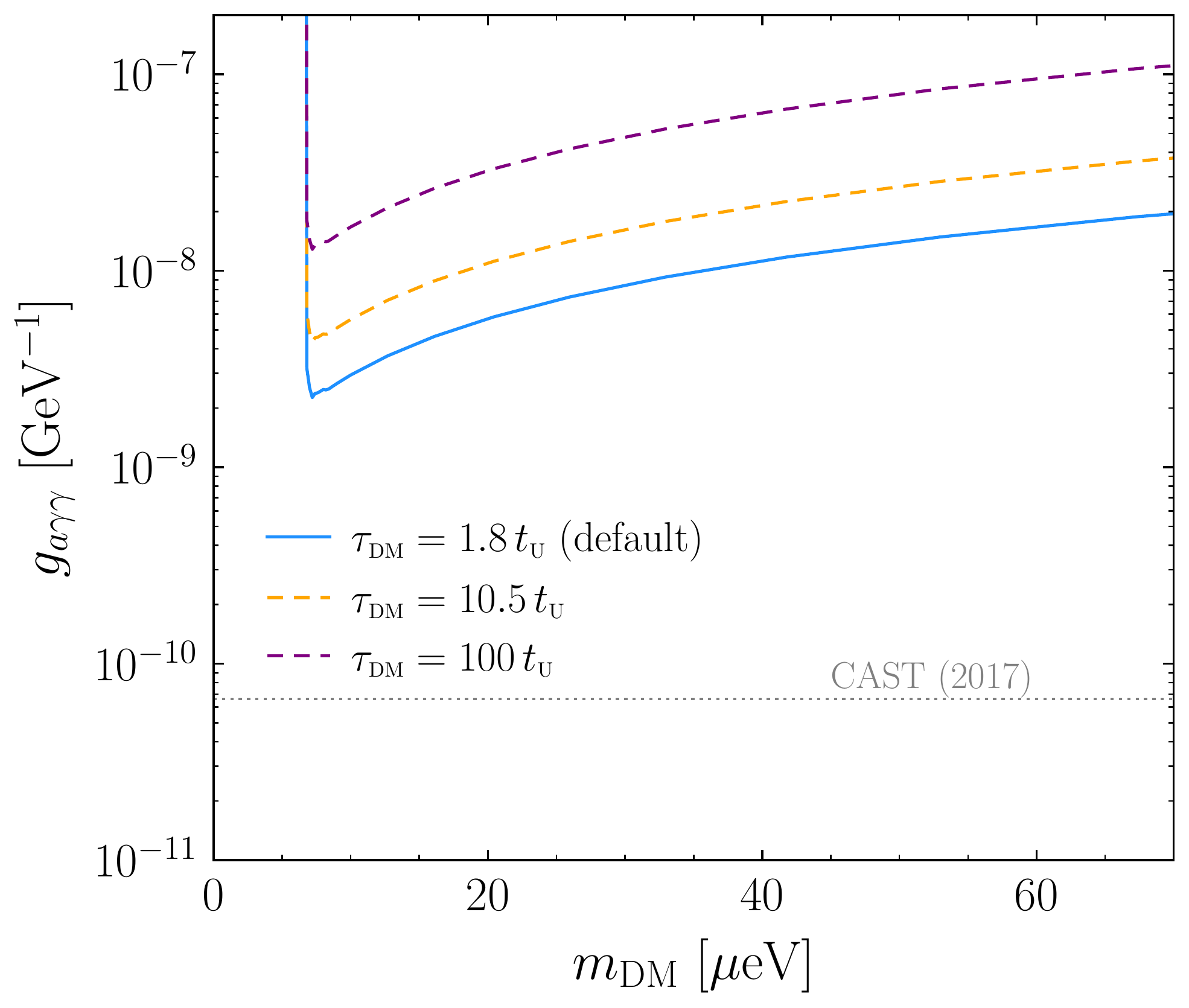}
\caption{Our fiducial result as in Fig.~\ref{fig:limits}, but for three different values of the dark-matter lifetime: $\tau = 1.8\,\tU$ (as adopted in the main text), but also $\tau=10.5\,\tU$ and $\tau=100\,\tU$.}
\label{fig:varytau}
\end{figure}

In the main text we chose to constrain our scenario for generating the \CaB by fixing the lifetime and then constraining the axion-photon coupling.
To this end, we used the 95\% C.L. established by DES on dark matter that decays to relativistic states in Ref.~\cite{PhysRevD.103.123528}, which required $\tau \geq 1.8\,\tU$.
That same work also studied the limit that can be obtained when their results are combined with measurements from the CMB, BAO, and also Type 1a supernovae.
Given these measurements are currently in tension, we chose not to use this as our default result, but the 95\% C.L. obtained from the combination requires $\tau \geq 10.5\,\tU$.
In terms of the fraction of dark matter that could have decayed to dark radiation, the combined limit corresponds to no more than 10\%, whereas the DES constraint alone is 72\%.
If in the future this was improved to 1\%, we would require $\tau \geq 100\,\tU$.
We show what our results from Fig.~\ref{fig:limits} in the main text would look like if we adopted these lifetimes in Fig.~\ref{fig:varytau}.

\end{document}